# Large-scale chromosome folding versus genomic DNA sequences: A discrete double Fourier transform technique

V. R. Chechetkin[a,b]* and V.V. Lobzin[c]

[a]*Engelhardt Institute of Molecular Biology of Russian Academy of Sciences, Vavilov str., 32, Moscow 119334, Russia*

[b]*Theoretical Department of Division for Perspective Investigations, Troitsk Institute of Innovation and Thermonuclear Investigations (TRINITI), Moscow, Troitsk District 108840, Russia*

[c]*School of Physics, University of Sydney, Sydney, NSW 2006, Australia*

___________________________

*Corresponding author. *E-mail addresses:* chechet@eimb.ru; vladimir_chechet@mail.ru

Tel.: +7 499 135 9895. Fax: +7 499 135 1405 (V.R. Chechetkin).

# ABSTRACT

Using state-of-the-art techniques combining imaging methods and high-throughput genomic mapping tools leaded to the significant progress in detailing chromosome architecture of various organisms. However, a gap still remains between the rapidly growing structural data on the chromosome folding and the large-scale genome organization. Could a part of information on the chromosome folding be obtained directly from underlying genomic DNA sequences abundantly stored in the databanks? To answer this question, we developed an original discrete double Fourier transform (DDFT). DDFT serves for the detection of large-scale genome regularities associated with domains/units at the different levels of hierarchical chromosome folding. The method is versatile and can be applied to both genomic DNA sequences and corresponding physico-chemical parameters such as base-pairing free energy. The latter characteristic is closely related to the replication and transcription and can also be used for the assessment of temperature or supercoiling effects on the chromosome folding. We tested the method on the genome of *Escherichia coli* K-12 and found good correspondence with the annotated domains/units established experimentally. As a brief illustration of further abilities of DDFT, the study of large-scale genome organization for bacteriophage PHIX174 and bacterium *Caulobacter crescentus* was also added. The combined experimental, modeling, and bioinformatic DDFT analysis should yield more complete knowledge on the chromosome architecture and genome organization.

# 1. Introduction

The genomic DNA of a living organism is compacted into a chromosome, an organized multilevel hierarchical structure. Combining imaging techniques and high-throughput genomic mapping tools, such as 3C, 4C, 5C, and Hi‑C, advanced the resolution of chromosome architecture and the information on chromatin interactions (Umbarger et al., 2011; Dekker et al., 2013; Hu et al., 2013; Nora et al., 2013; Bickmore, 2013; Pombo and Dillon, 2015; Sexton and Cavalli, 2015; Ramani et al., 2016). At each structural level, the chromosome folding consists of approximately repeating structural units. The quasi-regular chromosome folding is achieved by the cooperative binding of structural proteins to chromosomal DNA (Sandman et al., 1998; Lewin, 2000; Sumner, 2003; Marko, 2015). The structural proteins bind preferably to the specific motifs or to the stretches with biased nucleotide content. The positions of such motifs or stretches over the genome should be coordinated with the structural units of chromosome folding. Thus, the corresponding genome-scale periodicities in genomic DNA sequences ought to reflect the quasi-regularities of structural units in the chromosome folding. This means that a part of information on the large-scale chromosome folding can be obtained directly via studying genome-scale quasi-periodic patterns for the plethora of genomic DNA sequences stored in the databanks.

Approximate quasi-periodic regularities in genomic DNA sequences can be efficiently displayed via discrete Fourier transform (DFT) (McLachlan and Stewart, 1976; Chechetkin and Turygin, 1994, 1995a; Makeev and Tumanyan, 1996; Tiwari et al., 1997; Dodin et al., 2000; Lobzin and Chechetkin, 2000; Anastassiou, 2001; Yin and Yau, 2007; Paar et al., 2008; Wang and Stein, 2010; Marhon and Kremer 2011; Kravatskaya et al., 2011, 2013; and references therein). The quasi-periodic patterns can be detected by the peak harmonics in DFT spectrum. However, such approach is permissible only within the homogeneous spectral ranges without trends. This differs drastically from the case of the periodicities reflecting the approximate regularity of large-scale domains/units distributed over the genome and associated with the large-scale chromosome folding. The related peaks are squeezed in the range of low wave numbers, where the height of harmonics diminishes rapidly with the increase of wave numbers. Such a trend is generated by multifarious effects reflecting the mosaic patchiness of the genome. Any large-scale patchiness, like the 10–100 kb segments of different nucleotide composition, the long stretches of tandem repeats scattered over the genome, or the interleaving exon-intron stretches corresponding to the particular genes, will induce a trend in the related DFT spectrum at the low wave numbers (see a review by Lobzin and Chechetkin, 2000). To avoid the confusion, note that the periods of tandem repeats can be determined by standard DFT, whereas the variations in length of stretches with repeats and their positioning over the genome would generate a trend at the low wave numbers. On the other hand, the mosaic patchiness of the genome is related to the hierarchical chromosome folding. The relationship between the isochores and the chromatin organization is well-established (Vinogradov, 2003; Arhondakis et al., 2011; Frenkel et al., 2011; Bernardi, 2015; and references therein) as well as the structural role of tandem repeats and their involvement in the regulation of gene expression (Vogt, 1990; Lewin, 2000; Richard et al., 2008; Tung, 2011; and references therein). The strong interrelation between patchiness and periodicity effects for the large-scale domains/units associated with the large-scale chromosome folding leads to the simultaneous generation of both the strong trends and significant peaks in Fourier spectra. In addition, the large-scale periodic patterns have been subjected to the vigorous quasi-random modifications during molecular evolution and should be identified statistically.

The statistical assessment of peaks within spectral ranges with trends presents a difficult unsolved problem. Evidently, the random sequences of the same nucleotide composition cannot be used as a reference model for the assessment of large-scale regularities because they generate homogeneous Fourier spectra without trends. Simple de-trending in the spectral range at the low wave numbers may suppress, distort, or even ruin the quasi-periodic effects. The applicability of the random reshuffling of large-scale segments over the genome to the assessment of large-scale regularities also remains problematic. To solve the problem of the search for large-scale regularities in genomic DNA sequences on the background of the strong quasi-random evolutionary modifications, we developed an original discrete double Fourier transform (DDFT) and applied it to the study of the genome for *Escherichia coli* K-12. DDFT serves for the assessment of the number of large-scale domains/units at each level of hierarchical chromosome folding. The technique is specific to the type of nucleotides and their combinations and in this sense reflects the specificity of structural protein binding to chromosomal DNA. The comparison of the annotated domains/units in the chromosome of *E. coli* proved to be in good agreement with the large-scale regularities in the genomic DNA sequences detected by DDFT.

## 2. Theory and methods

### 2.1. Discrete Fourier transform (DFT)

In this and next sections we follow the methods developed previously (Chechetkin and Turygin, 1994, 1995a; Lobzin and Chechetkin, 2000). Fourier harmonics corresponding to nucleotides of type $\alpha \in$ (A, C, G, T) in a genomic sequence of length $M$ are calculated as

$$\rho_\alpha(q_n) = M^{-1/2} \sum_{m=1}^{M} \rho_{m,\alpha} e^{-iq_n m}, \quad q_n = 2\pi n/M, \quad n = 0, 1, ..., M-1 \qquad (1)$$

Here $\rho_{m,\alpha}$ indicates the position occupied by the nucleotide of type $\alpha$; $\rho_{m,\alpha} = 1$ if the nucleotide of type $\alpha$ occupies the $m$-th site and 0 otherwise. The amplitudes of Fourier harmonics (or structure factors) are expressed as

$$F_{\alpha\alpha}(q_n) = \rho_\alpha(q_n) \rho_\alpha^*(q_n) \qquad (2)$$

where the asterisk denotes the complex conjugation. The harmonics with $n = 0$ depend only on the nucleotide composition. They do not contain structural information and will be discarded. Due to the symmetry of structure factors,
$$F_{\alpha\alpha}(q_n) = F_{\alpha\alpha}(2\pi - q_n) \qquad (3)$$

Fourier spectrum can be restricted from $n = 1$ to
$$N = [M/2] \qquad (4)$$

where the brackets denote the integer part of the quotient. The structure factors will always be normalized relative to the mean spectral values, which are determined by the exact sum rules,

$$f_{\alpha\alpha}(q_n) = F_{\alpha\alpha}(q_n)/\overline{F}_{\alpha\alpha}; \quad \overline{F}_{\alpha\alpha} = N_\alpha(M - N_\alpha)/M(M-1) \qquad (5)$$

where $N_\alpha$ is the total number of nucleotides of type $\alpha$ in a sequence of length $M$.
In addition to the normalized structure factors, their various sums will be used below,
$$S_{\alpha\beta}(q_n) = f_{\alpha\alpha}(q_n) + f_{\beta\beta}(q_n); \alpha \neq \beta \qquad (6)$$

$$S_{\alpha\beta\gamma}(q_n) = f_{\alpha\alpha}(q_n) + f_{\beta\beta}(q_n) + f_{\gamma\gamma}(q_n); \alpha \neq \beta \neq \gamma \qquad (7)$$

$$S_4(q_n) = f_{AA}(q_n) + f_{CC}(q_n) + f_{GG}(q_n) + f_{TT}(q_n) \qquad (8)$$

Note that the normalized structure factors contain information about the regularity of nucleotide positions and do not depend on the nucleotide content. For the random sequences, normalized structure factors for all nucleotides obey identical Rayleigh statistics. Therefore, the weighing of additives in the sums (6)–(8) by the nucleotide frequencies would be incorrect. Using sums (6)–(8) is more convenient than the grouping of the corresponding nucleotides directly in the genomic sequence. In particular, the binary grouping like A–non-A, where non-A = (T, C, G), directly in the genomic sequence would provide the same Fourier spectra $f_{AA}(q_n) = f_{non-A, non-A}(q_n)$, whereas the combination $f_{CC}(q_n) + f_{GG}(q_n) + f_{TT}(q_n)$ contains the information different from that in $f_{AA}(q_n)$.

*2.2. Correlation functions*

The study of circular correlation functions for the nucleotides of type α,

$$K_{\alpha\alpha}(m_0) = M^{-1} \sum_{m=1}^{M} \rho^c_{m,\alpha} \rho^c_{m+m_0,\alpha}, \quad m_0 = 0, 1, ..., M-1 \tag{9}$$

$$\rho^c_{m,\alpha} = \begin{cases} \rho_{m,\alpha}, & \text{if } 1 \le m \le M \\ \rho_{m-M,\alpha}, & \text{if } M+1 \le m \le 2M-1 \end{cases} \tag{10}$$

complements the analysis of regularities in the genomic nucleotide sequences via DFT. The correlation functions and the structure factors are not independent and are related by the Wiener-Khinchin relationship,

$$K_{\alpha\alpha}(m_0) = M^{-1} \sum_{n=0}^{M-1} F_{\alpha\alpha}(q_n) \exp(-i q_n m_0) \tag{11}$$

The correlation functions are symmetrical relative to the middle,

$$K_{\alpha\alpha}(m_0) = K_{\alpha\alpha}(M - m_0) \tag{12}$$

The corresponding mean value is given by

$$\overline{K}_{\alpha\alpha} = \frac{1}{M-1} \sum_{m_0=1}^{M-1} K_{\alpha\alpha}(m_0) = \frac{N_\alpha(N_\alpha - 1)}{M(M-1)} \tag{13}$$

The normalized deviations,

$$\kappa_{\alpha\alpha}(m_0) = \left(K_{\alpha\alpha}(m_0) - \overline{K}_{\alpha\alpha}\right) / <\Delta K^2_{\alpha\alpha}>^{1/2}_{random} \tag{14}$$

where

$$<\Delta K^2_{\alpha\alpha}>_{random} = \overline{F}^2_{\alpha\alpha} / M, \quad \overline{F}_{\alpha\alpha} = N_\alpha(M - N_\alpha) / M(M-1) \tag{15}$$

are Gaussian for the random sequences. Similarly to the sums (6)–(8), it useful to introduce the combinations,

$$Q_{\alpha\beta}(m_0) = \left(\kappa_{\alpha\alpha}(m_0) + \kappa_{\beta\beta}(m_0)\right) / \sqrt{2}; \quad \alpha \ne \beta \tag{16}$$

$$Q_{\alpha\beta\gamma}(m_0) = \left(\kappa_{\alpha\alpha}(m_0) + \kappa_{\beta\beta}(m_0) + \kappa_{\gamma\gamma}(m_0)\right) / \sqrt{3}; \quad \alpha \ne \beta \ne \gamma \tag{17}$$

$$Q_4(m_0) = \left(\kappa_{AA}(m_0) + \kappa_{TT}(m_0) + \kappa_{CC}(m_0) + \kappa_{GG}(m_0)\right) / 2 \tag{18}$$

which are also approximately Gaussian for the random sequences. The sums (16)–(18) are sensitive to the mutual signs of different deviations. If the signs of different deviations are not interesting, the significance of deviations for the different combinations of nucleotides can be assessed by the respective $\chi^2$-criteria.

*2.3. Discrete double Fourier transform (DDFT)*

The method presented in this section is original and is not described in the vast literature on Fourier transform. Therefore, first, we should explain why DDFT? As has been proved earlier (Chechetkin and Turygin, 1995a; Lobzin and Chechetkin, 2000; Paar et al., 2008), the periodic patterns of period $p = M/n$ generate a set of equidistant harmonics with wave numbers $n$, $2n$, ..., $k_{max}n \leq N$. The proof of the statement is simple. If the nucleotides occupy $p$-periodic positions, the corresponding ratios $mn/M$ will be equal to integers at the wave number $n = M/p$, the exponents in Eq. (1) will be phased, $e^{-iq_n m} = e^{-i2\pi mn/M} = e^{-i2\pi m/p} = 1$, and produce a peak in Fourier spectrum. The similar succession of statements will also be valid for the wave numbers $2n$, ..., $k_{max}n \leq N$, that means a series of equidistant peaks. Such series of peaks can be easily obtained for a sequence of tandem repeats. This is the reason why using wave numbers $n$ in the presentation of Fourier spectra is more preferable than using periods $p$, because the equidistant series may refer to the same period. The variations in periods of patterns induce the deviations in equidistance and may partially suppress the harmonics with the higher wave numbers $kn$ (Lobzin and Chechetkin, 2000). The random point mutations and indels in the periodic patterns may strongly distort this idealized picture and the periodicities become hidden. Generally, hidden periodicities can be detected either by statistically significant singular high peaks and/or by the sums of equidistant harmonics (Chechetkin and Turygin, 1995a; Lobzin and Chechetkin, 2000).

As discussed in Introduction, the analysis of genome-scale periodicities in DNA sequences potentially associated with the large-scale chromosome folding is seriously hampered by the interrelation of inhomogeneity and periodicity effects. For example, distribution of large-scale AT- or GC-rich stretches over the genome may be either quasi-periodic or not, yet in both cases the high harmonics will be induced in DFT spectrum at the low wave numbers. The mosaic patchiness of the genome or finite-range correlations in DNA sequences also generate high peaks and a trend in Fourier spectrum at the low wave numbers (Lobzin and Chechetkin, 2000; and references therein). In the ranges with trends, the high peaks at the low wave numbers cannot serve for the correct assessment of the periodicity effects due to the strong contribution of the other effects. The additional cross-check is needed to filter out the numerous false positives attributed to the large-scale quasi-periodic patterns in DFT spectrum. Fortunately, there is an important distinction between the inhomogeneity or finite-range correlations and periodicity effects in Fourier spectra. The inhomogeneity or finite-range correlations do not induce the equidistant series of harmonics, whereas the periodicity does. The simultaneous observation of a high harmonic at a low wave number and the corresponding equidistant series of harmonics indicates the "true periodicity" at the large scales, the feature related to equidistant series being unique and more important statistically. The set of equidistant harmonics induced in DFT spectrum by a genome-scale periodicity is large enough to be detected by the iteration of Fourier transform. For the practical purposes, one iteration or the double Fourier transform is quite suitable. To differentiate genome-scale periodicities from short-range ones comparable with the helix pitch of double-stranded DNA (dsDNA), such long-range periodicities will be termed superperiods.

Previously, it was suggested to identify hidden periodicities via the sums of equidistant harmonics surrounded by variable windows (Chechetkin and Turygin, 1995a; Chechetkin and Lobzin, 1998; Lobzin and Chechetkin, 2000). This technique is inconvenient because the number of equidistant harmonics and the width of windows vary in the different sums and affect the relevant statistical criteria for the assessment of significance of observed periodicity. DDFT is free from these drawbacks and provides a unified approach to the study of hidden equidistance in DFT spectra. Multiple harmonics in a set with hidden equidistance associated with large-scale quasi-periodic patterns in primary DFT spectrum are reduced to few harmonics by DDFT that makes the analysis much more efficient.

The extrapolation of equidistance related to the genome-scale periodicities onto the range of short-distance periodicities in Fourier spectra needs an additional study. One of the basic short-distance periodicities refers to helix pitch of dsDNA and is about $p = 10.5$. The A-tracts, defined as the sequences $A_nT_m$ phased with the helix pitch, can induce the intrinsic curvature of dsDNA and facilitate the chromosome folding (Trifonov and Sussman, 1980; Hagerman, 1990; Crothers et al., 1990; Trifonov, 1998; Hizver et al., 2001; Tolstorukov et al., 2005; and references therein).The alternative motifs, G-tracts ($G_nC_m$), are related to DNA bending into the major groove, especially in the presence of divalent cations, and also affect the folding (Milton et al., 1990; Brukner et al., 1993; Goodsell et al., 1993; Hud and Plavec, 2003). The intensity of periodicity $p = 10.5$ and the variations around $p = 10.5$ associated with DNA curvature or supercoiling in the vicinity of promoter regions may strongly affect the gene expression (Pérez-Martín and De Lorenzo, 1997; Jáuregui et al., 2003; Schieg and Herzel, 2004; Klaiman et al., 2009; Kravatskaya et al., 2011, 2013). The other basic periodicity $p = 3$ is inherent to the protein-coding regions (see, e.g. Tiwari et al., 1997; Trifonov, 1998; Lobzin and Chechetkin, 2000; Yin and Yau, 2007; Wang and Stein, 2010; Marhon and Kremer, 2011; and references therein). To assess the effects of short-range periodicities on the hidden equidistance related to the genome-scale periodicities, we introduced the variable cut-off $N_c$ for the right part of Fourier spectrum.

In the noisy DFT spectrum, the regularities can be detected if the number of significantly high equidistant harmonics exceeds $\sqrt{N_c}$. Therefore, the cut-off $N_c$ should satisfy the inequality

$$N_c/n \gg \sqrt{N_c} \quad \text{or} \quad \sqrt{N_c} \gg n \tag{19}$$

where $n$ corresponds to the wave number associated with the genome-scale periodicity under study. The inequality (19) means that the total number of equidistant harmonics generated by the harmonic with wave number $n$ in the range up to cut-off $N_c$ must exceed the detection threshold $\sqrt{N_c}$. It also provides the restriction on the wave numbers $n$ for which the equidistance can be resolved with DDFT at fixed cut-off $N_c$.

The harmonics in DDFT are defined as

$$\Phi_\alpha(\tilde{q}_{n'}) = (N_c - 1)^{-1/2} \sum_{n=2}^{N_c} f_{\alpha\alpha}(q_n) e^{-i\tilde{q}_{n'}n}, \quad \tilde{q}_{n'} = 2\pi n'/(N_c - 1), \quad n' = 0, 1, ..., N_c - 2 \tag{20}$$

where $N_c$ is cut-off and $f_{\alpha\alpha}(q_n)$ are the normalized structure factors (5). The structure factor harmonic with $n = 1$ does not induce equidistant series and is discarded from the sum. The hidden equidistance is assessed by the amplitudes

$$F_{\alpha\alpha,II}(\tilde{q}_{n'}) = \Phi_\alpha(\tilde{q}_{n'})\Phi_\alpha^*(\tilde{q}_{n'}) \tag{21}$$

Again, we are interested in harmonics with non-zero wave numbers $n'$ and can restrict the spectrum to the left half from $n' = 1$ to

$$N' = [(N_c - 1)/2] \tag{22}$$

due to the symmetry relationship (cf. Eqs. (3) and (4)). For the complete genome, the right spectral boundary in DDFT spectrum is $N' = [(N - 1)/2]$, where $N$ is defined by Eq. (4) and the brackets in Eq. (22) and in this definition denote the integer part of the quotient. The normalization of the amplitudes (21) is performed as

$$f_{\alpha\alpha, II}(\tilde{q}_{n'}) = F_{\alpha\alpha, II}(\tilde{q}_{n'}) / \overline{F}_{\alpha\alpha, II} \tag{23}$$

$$\overline{F}_{\alpha\alpha, II} = \frac{1}{N'} \sum_{n'=1}^{N'} F_{\alpha\alpha, II}(\tilde{q}_{n'}) \tag{24}$$

DDFT for the sums (6)–(8) is defined in lines with Eqs. (20)–(24). In the latter case the normalized structure factors $f_{\alpha\alpha}(q_n)$ in (20) should be replaced by the corresponding sums $S(q_n)$. The order of operations (first, summation of the counterpart structure factors, then, application of DDFT to the sums) enhances the sensitivity of hidden equidistance detection.

The number of superperiods associated potentially with the number of large-scale quasi-regular structural units in hierarchical chromosome folding can be assessed by the wave number $n'$ for the singular high amplitude $f_{\alpha\alpha, II}(\tilde{q}_{n'})$ as

$$N_{\text{superperiods}} \equiv N_{sp} = (N_c - 1)/n' \tag{25}$$

whereas their periods in base pairs are given by

$$p_{\text{superperiods}} = M / N_{\text{superperiods}} = Mn'/(N_c - 1) \tag{26}$$

The significance of superperiods detected by DDFT should generally be assessed either by the singular peak amplitudes in DDFT spectrum or by the series of equidistant amplitudes with $k'n', k' = 1, ..., k'_{\max}; k'_{\max} n' \leq N'$, where $N'$ is defined by Eq. (22) (cf. Chechetkin and Turygin, 1995a; Lobzin and Chechetkin, 2000). The variations in equidistant series in primary DFT spectrum would result in the variable equidistance for the related DDFT spectrum and may also lead to the suppression of the amplitudes with the higher wave numbers $k'n'$ (Lobzin and Chechetkin, 2000).

*2.4. Base-pairing free energy*

DNA thermodynamics affects chromosome architecture and topology (Travers and Muskhelishvili, 2013). Unzipping dsDNA during replication or transcription is initiated by localized unwinding of specific DNA sequences recognized by specific proteins and depends on base-pairing free energy (Lewin, 2000). Relative stability of dsDNA has been used to predict the promoter regions in bacterial genomes responsible for the transcription initiation (Rangannan and Bansal, 2010). The hierarchical folding of the *E. coli* nucleoid (Dorman, 2013) proved to be correlated with the hierarchy of transcriptional patterns (Jeong et al., 2004). Therefore, we also studied large-scale distribution of base-pairing free energy over the genome as potential indicator of domains/units in the chromosome folding. The free energy for a DNA stretch was additively calculated by the partial energies of corresponding tiling dinucleotides, e.g., 5'-ATGGCC-3' = AT + TG + GG + GC + CC. If the sequence 5'-ATGGCC-3' would be circular, the additive corresponding to the terminal pair CA should append the sum. The free energy parameters for the particular dinucleotides were taken from the paper by SantaLucia and Hicks (2004).

DFT for base-pairing free energy is defined as

$$\rho_{FE}(q_n) = M'^{-1/2} \sum_{m'=1}^{M'} G_{m'} e^{-iq_n m'}, \quad q_n = 2\pi n / M', \quad n = 0, 1, ..., M'-1, \tag{27}$$

where $m'$ numerates the tiling dinucleotides over the genome, $M' = M - 1$ for the linear genome of length $M$ and $M' = M$ for the circular genome, and $G_{m'}$ corresponds to free energy for the $m'$-th dinucleotide. All other definitions are in lines with Eqs. (2)–(4). The expression for the mean structure factor in DFT spectrum is

$$\overline{F}_{FE} = \frac{1}{M'-1}\sum_{n=1}^{M'-1} F_{FE}(q_n) = \frac{1}{M'-1}\sum_{m=1}^{M'}(G_m - \overline{G})^2 = \sigma^2(G) \qquad (28)$$

and should be used for the normalization similarly to Eq. (5). DDFT procedure for free energy corresponds to that of described in Section 2.3.

The circular correlation function for free energy and the related normalized deviations are defined similarly to Eqs. (9)–(14). The corresponding mean values are given by

$$\overline{K}_{FE} = \frac{1}{M'-1}\sum_{m_0=1}^{M'-1} K_{FE}(m_0) = \left(M'\overline{G}^2 - \overline{G}^2\right)/(M'-1) \qquad (29)$$

$$\overline{G} = \frac{1}{M'}\sum_{m=1}^{M'} G_m \qquad (30)$$

$$\overline{G^2} = \frac{1}{M'}\sum_{m=1}^{M'} G_m^2 \qquad (31)$$

$$<\Delta K_{FE}^2>_{random} = \overline{F}_{FE}^2 / M' \qquad (32)$$

As the correlations for free energy in random nucleotide sequences correspond to Markov chains with one-step memory, it would be more correct to separate out the correlations at $m_0 = 1$ and redefine the mean (29) and the dispersion (32) only for the components $2 \le m_0 \le [M/2]$. This is not, however, essential for our subsequent analysis, because the difference between the mean (29) and the dispersion (32) and the redefined values is small and can be neglected.

In the definition of free energy,

$$G = H - TS \qquad (33)$$

the energy $H$ and the entropy $S$ are assumed to be temperature independent. This implies relatively narrow temperature range around 37°C. In a broader range, the temperature dependence of the parameters $H$ and $S$ should be taken into account as well.

*2.5. Statistical criteria*

The statistical significance of the heights of structure factors (5) in the whole spectrum or in a range of the spectrum should be assessed by the extreme value statistics. The same concerns the swings of normalized deviations for correlation functions (14). Let the variables in a set $\{y_i\}_{i=1}^{N}$ be identically distributed and let the probability that one variable will exceed a threshold $y'$ be $\Pr(y > y')$. Then, the probability that at least one from $N$ variables $\{y_i\}_{i=1}^{N}$ exceeds $y'$ is given by (Johnson and Leone, 1977),

$$\Pr(y > y'; N) = 1 - (1 - \Pr(y > y'))^N \qquad (34)$$

We will use for the probability $\Pr(y > y'; N)$ the standard statistical threshold $\Pr_{thr} = 0.05$. The Bonferroni approximation yields a good estimate for the corresponding threshold $y'$,

$$\Pr(y > y_{thr}) \approx \Pr_{thr}/N \qquad (35)$$

The relationship (35) means that the probability threshold for the multiple $N$ trials diminishes about $N$ times in comparison with the threshold for one trial.

The structure factors (5) for random sequences obey the Rayleigh distribution (Chechetkin and Turygin, 1994; Lobzin and Chechetkin, 2000),

$$\Pr(f > f') = \exp(-f') \tag{36}$$

For the sums (6)–(8) the corresponding distributions are

$$P(S > S') = e^{-S'} \sum_{k=1}^{r} \frac{S'^{k-1}}{(k-1)!}; \; r = 2, 3, 4 \tag{37}$$

The Rayleigh distribution provides also a reasonable approximation for the normalized amplitudes (23) in random DDFT spectra. The same concerns DDFT counterparts for the variables (6)–(8) (Fig. 1).

The probability that the modulus of normalized deviations (14) for random sequences exceeds $|\kappa'|$ is Gaussian,

$$\Pr(|\kappa| > |\kappa'|) = \sqrt{\frac{2}{\pi}} \int_{|\kappa'|}^{\infty} dy \exp(-y^2/2) \tag{38}$$

The same is true for the sums (16)–(18). The Gaussian extreme value thresholds for $N$ about $10^5$–$10^6$ are higher than the thresholds 3–4 used often by convention. The probabilities (36)–(38) should be substituted into Eq. (34) at the study of respective problem.

*2.6. Preprocessing of Fourier spectra*

The mean spectral values for the structure factors (2) are determined by the exact sum rules (Chechetkin and Turygin, 1994; Lobzin and Chechetkin, 2000; Chechetkin, 2011). They are independent of the regularity of nucleotide distribution over the genome and depend only on the nucleotide content (see Eq. (5)). As the mean spectral value is fixed, the clusters of high harmonics in separate parts of Fourier spectrum diminish the heights of harmonics in other parts of the spectrum and lead to the statistical underestimation of the less pronounced periodicities. The rigorous theory for the treatment of this effect should be based on the conditional probability and is rather tedious. Here, we describe a simplified procedure aimed at the solution of this problem. Let the spectral range corresponding to the structure factors with wave numbers $n \in (n_1, n_2)$ be approximately homogeneous. Calculate the local mean value within this range,

$$\bar{f}_{n_1 n_2} = \frac{1}{n_2 - n_1 + 1} \sum_{n=n_1}^{n_2} f(q_n) \tag{39}$$

and renormalize the structure factors (5) according to

$$f_{renorm}(q_n) = f(q_n) / \bar{f}_{n_1 n_2}; \; n \in (n_1, n_2) \tag{40}$$

How does such renormalization affect the statistical criteria? We are interested in the statistical properties of renormalized structure factors

$$f_{renorm} = rf / S_r; \; S_r = f_1 + \ldots + f_r \tag{41}$$

All variables $f_i$ are implied to be identically distributed by the Rayleigh distribution. Then, the corresponding probability density for $f_{renorm}$ is defined as

$$p(f_{renorm}) = \int_0^\infty df \int_0^\infty dS \, \delta(f_{renorm} - rf/S) \, e^{-f} \, \frac{S^{r-1}}{(r-1)!} e^{-S} = (1 + f_{renorm}/r)^{-(r+1)} \quad (42)$$

At the limit of large $r$, $p(f_{renorm})$ can be approximated by the Rayleigh probability density, $p(f_{renorm}) \approx e^{-f_{renorm}}$. This means that the Rayleigh distribution is robust relative to the local spectral renormalization of the structure factors within a chosen spectral range, if such spectral range contains many harmonics. After renormalization within chosen range(s), the corresponding renormalization within the other ranges should also be performed. The complete renormalization preserves the sum rules.

The other problem concerns the inhomogeneous spectral ranges with trends, typically, at the low wave numbers. In this range, the heights of harmonics lower rapidly with the increase of wave numbers. As was argued in Section 2.3, the de-trending for DFT spectra in the range at the low wave numbers is undesirable at the study of genome-scale periodicities and the raw data are preferable for the preliminary assessment of genome-scale regularities. The corresponding de-trending for DDFT spectra is less restrictive, because the peaks related to hidden equidistance are mainly in the right part of DDFT spectra at the relatively high wave numbers. Let the range with a trend be restricted by harmonics having the wave numbers $n_L$ and $n_R$ and let trend for the heights of harmonics in this range be approximated by a polynomial $a_0 + a_1 n + ... + a_k n^k$, where $n$ is harmonic number and $\{a_i\}_{i=0}^k$ is the set of coefficients. The de-trending factor is defined as

$$k_{de-trend}(n) = \left(a_0 + a_1 n + ... + a_k n^k\right) / \left(a_0 + a_1 n_R + ... + a_k n_R^k\right); \, n \in (n_L, n_R) \quad (43)$$

while the amplitudes of harmonics should be de-trended as

$$f_{de-trend}(q_n) = f(q_n) / k_{de-trend}(n); \, n \in (n_L, n_R) \quad (44)$$

If needed, the non-polynomial scheme can be used for de-trending.

At the end of this section, we will comment on Fourier transform for free energy. The overlapping produces correlations between consecutive dinucleotides even in random sequences. The corresponding random models for nucleotides and dinucleotides refer to Markov chains with zero and one-step memory. The strong correlations at $m_0 = 1$ for free energy generate a wide trend about $M/2$ within the range of wave numbers in DFT spectrum. This trend will be squeezed at the low wave numbers after DDFT. The resulting distribution for the whole DDFT spectrum corresponding to free energy profile in a random sequence would be different from the Rayleigh distribution, whereas the local normalization (40) within DDFT spectral ranges without trends restores the Rayleigh distribution for harmonics in such ranges. This adds arguments in favor of preprocessing of DDFT spectrum rather than DFT one. Note that all statements on Fourier transform for free energy were verified by test simulations with random nucleotide sequences.

## 3. Chromosome organization of *Escherichia coli*

As an example of the application of our methods, we chose the well-studied chromosome organization for the model Gram-negative bacterium *E. coli*. The genome of *E. coli* strain K-12 substrain MG1655 (GenBank assembly accession GCA_000005845.2) contains 4,641,652 bp and codes for more than 4,200 genes. We are interested in the large-scale (on the genome size) domains/structural units in the *E. coli* chromosome (reviewed by Dorman, 2013; Messerschmidt and Waldminghaus, 2014). Genetically, the *E. coli* chromosome can be divided by the right (clockwise) and left (anticlockwise) replichores and, further, by four macrodomains, Ori, Ter, Left, and Right, and two so-called non-structured regions, NS-left and NS-right. These macrodomains are unlikely to undergo recombination with each other because such rearrangements affect negatively the survival of the bacterium (Valens et al., 2004). The families of proteins with macrodomain-specific DNA-binding properties have been identified (Dame et al., 2011).

The folding of the *E. coli* nucleoid is hierarchical and dynamic (depends on the cell cycle). In rapidly growing bacteria, the first three levels of the hierarchical nucleoid folding contain (i) a writhed structure (resembling a double-twisted thick ring); (ii) solenoid and plectonemic helices of approximately 117 kb per turn; and (iii) 10–12 microdomain loops within 117 kb turns. The domains/units in the writhed structure (level (i)) can be determined only approximately. If at least part of adjacent solenoid and plectonemic turns (level (ii)) is regularly alternating, the corresponding turns can be united into the longer turns of doubled pitch.

The folding of the *E. coli* nucleoid should ensure the efficient regulation of gene expression. Jeong et al. (2004) studied the transcription in the genome of *E. coli* K-12 as a function of the position of genes on the chromosome and revealed the spatial variations in transcriptional activity. The hierarchical transcriptional patterns can be classified into long-range, over 600-800 kb; medium-range, over 100-125 kb; and short-range, of up to 16 kb, categories. The characteristic lengths of the transcriptional patterns appear to be close to that corresponding to the genetic macrodomains and the structural units in the *E. coli* nucleoid (summarized in Table 1).

**Table 1**

Characteristic large-scale domains/units in the chromosome of *E. coli*

| Domains/units | Total number of domains/units | Characteristic length |
|---|---|---|
| Genetics[a] | | |
| Right (clockwise) and left (anticlockwise) replichores | 2 | 2.3 Mb |
| Ori, NS-right, Right, Ter, Left, NS-left | 6 | 770 kb |
| Nucleoid folding[a] | | |
| A writhed structure | 6 | 770 kb |
| A solenoid and a plectoneme, both of periodicity 117 kb | 40 | 117 kb |
| 10–12 microdomains within 117 kb units | 400–480 | 10–12 kb |
| Gene expression[b] | | |
| Long-range patterns | 6–8 | 600–800 kb |
| Medium-range patterns | 37–46 | 100–125 kb |
| Short-range patterns | 290 | up to 16 kb |

[a]Dorman (2013); [b]Jeong et al. (2004).

Could the numbers of relevant large-scale domains/units be assessed directly via underlying genomic DNA sequences? We try to answer this question in the next section. We retain the terminology by Jeong et al. (2004) for the hidden regularities in the *E. coli* genome potentially associated with the large-scale domains/units and classify the corresponding superperiods also into long, medium, and short categories.

## 4. Results

### 4.1. Long and medium superperiods in the E. coli genome

The numbers of domains/units at the consecutive levels of hierarchical chromosome organization differ by the factor of 5–10 (Chechetkin, 2013; and references therein). Therefore, we define the categories of superperiods associated with the different domains/units as: long superperiods, range of 2–12 superperiods; medium superperiods, range 20–100; and short superperiods, range 190–900. Each range was chosen to be broader than the counterpart ranges listed in Table 1 for the *E. coli* chromosome to detect the difference between hierarchical levels of chromosome organization and to verify the non-random origin of detected superperiods. The detailed information on the primary DFT spectra for the combinations (5)–(8) and for the base-pairing free energy profiles at the different temperatures can be found in Supplementary file S1.

Throughout Section 4, the number of superperiods identified via the highest harmonics in DDFT spectra should be compared with their counterparts listed in the middle column of Table 1 and corresponding to the structural and biological features of *E. coli*. To assess the robustness of detected superperiods, we used DDFT for all 15 variables (5)–(8) and the base-pairing free energy profile at 37°C. The choice of particular nucleotide combinations may reflect the specificity of binding motifs for the architectural proteins. The resulting distributions of the detected superperiods should be compared against the analogous distributions for the random spectra. In the hierarchical chromosome folding, the units at the higher levels of folding are composed of the units at the lower levels. The differentiation in the levels of chromosome folding results in the separation of the related superperiods which can be resolved by DDFT. In some cases, the superperiods may be attributed to the coordinated modifications of the shorter superperiods at the same level of folding (see examples with period-doubling below and the example in Section 4.5).

DDFT spectrum and the spectral ranges corresponding to the long and medium superperiods obtained for DDFT of the sum $S_4(q_n)$ (Eq. (8)) are shown in Fig. 2. Both spectral ranges were renormalized on the local means as described in Section 2.6. The numbers of superperiods detected by the peak harmonics are marked by arrows and reveal good accordance with the numbers of the counterpart domains/units in Table 1. The very high peaks at $n = 2$, 3, and 6 in the corresponding primary DFT spectrum (Supplementary file S1) indicate that these superperiods should be treated separately rather than as the approximate equidistant set for $N_{sp} = 6.3$ (as discussed at the end of Section 2.3). The medium superperiods with $N_{sp} = 57.8$ and 28.9 (Fig. 2B) can be related to the partial alternation of the solenoidal and plectonemic turns. The harmonic with $n = 29$ is also among the significant harmonics in the primary DFT spectrum, whereas the harmonic with $n = 58$ is absent (note, however, the significant harmonic with $n = 57$). The correspondence between the medium superperiods with $N_{sp} = 43.9$ and the highly significant harmonic with $n = 44$ in DFT spectrum is fulfilled.

The short superperiods is better to study separately by modifying the main scheme (presented in the next section). In this section we restrict ourselves to the long and medium superperiods. The superperiods detected by the three highest harmonics in relevant DDFT spectral ranges obtained for the variables (5)–(8) and for the base-pairing free energy profile at 37°C are summarized in Fig. 3 and Supplementary file S2. The distributions of superperiods in Figs. 3A and 3B were compared with the related coarse-grained distributions for the random sequences. The Fourier harmonics with the different wave numbers are uncorrelated for the random sequences (both in DFT and in DDFT spectra). Therefore, the frequency to detect a peak harmonic is proportional to the width of a chosen spectral range in wave numbers. The coarse-graining for the long superperiods was performed as follows. The interval for the terminal superperiods 2 and 12 was chosen to be 0.5 and was one-sided, whereas all other superperiods were surrounded by the intervals 0.5 around integer superperiods. Then, the fractions of superperiods within such intervals were calculated for the *E. coli* genome and compared with related fractions calculated for the random sequences (Fig. 3C). The corresponding dispersions may be assessed by the binomial distributions, $\sigma^{(r)2} = \phi_{random}^{(r)}(1-\phi_{random}^{(r)})/N_{total}$ (where $\phi_{random}^{(r)}$ is the fraction of detected peaks within the chosen range $r$ for the random sequences and $N_{total} = 48$ is the total number of ranked peaks for all variables (5)–(8) and free energy), while the detected difference $\phi_{E.coli}^{(r)} - \phi_{random}^{(r)}$ may be assessed against $\sigma^{(r)}$. The similar coarse-graining with the interval 5 was applied to the medium superperiods (Fig. 3D).

The distinct gap beyond $N_{sp} = 6$ for the long superperiods (Figs. 3A and 3C) indicates clearly the hierarchical mode of chromosome organization. The detection of the separate superperiods with $N_{sp} \approx 2$, 3, and 6 for the particular combinations (5)–(8) provides additional arguments in favour of their independence. The robust reproducibility of superperiods with the same wave numbers $n'$ for the several combinations (5)–(8) enhances their statistical significance. In particular, the superperiods with $N_{sp} = 3.0$ ($n' = 763823$) were robustly detected for $f_{AA}(q_n) + f_{GG}(q_n)$, all triple combinations (7) except $f_{TT}(q_n) + f_{CC}(q_n) + f_{GG}(q_n)$ (the superperiods close to 3.0 being detected for this combination as well), and $S_4(q_n)$ (Supplementary file S2). The scattering in $N_{sp}$ for the different combinations of nucleotides reflects the dynamic nature of nucleoid folding and the participation of the different families of the architectural proteins in the folding (different specificity of protein-DNA binding) and is partially due to the quasi-random modifications in genomic DNA sequences during molecular evolution. Although the harmonic with $n = 4$ is absent among significant harmonics in the primary DFT spectra, the bias in the fraction of superperiods with $N_{sp} \approx 4$ (Fig. 3C) indicates their significance. The resulting nucleoid conformation depends on the supercoiling. The level of DNA negative supercoiling in log phase is higher in comparison with lag and stationary phases and affects nucleoid topology (Dorman, 2013). The distribution of supercoiling over the genome varies during the cell cycle (Lal et al., 2016). The superperiods with $N_{sp} \approx 4$ might indicate an intermediate once-twisted thick ring (or figure of eight) conformation for the writhed structure. The transitions between once- and double-twisted conformations during cell cycle look to be quite plausible.

For the medium superperiods, the solenoidal and plectonemic 117 kb-turns vary with time and along the chromosome. The unification of the solenoidal and plectonemic turns may be responsible for the shift toward lower number of superperiods from the expected value about $N_{sp} \approx 40$ for the medium superperiods (Table 1 and Figs. 3B, 3D). The significant counterparts in the primary DFT spectra can be found for nearly all medium superperiods detected by DDFT.

DDFT spectra were also studied at the different cut-offs $N_c = 600$; 1,200; 10,000; and 100,000 in the primary DFT spectra (Section 2.3). We found that the characteristic superperiods related to the peak harmonics are mainly retained, however, all DDFT spectra with cut-offs were distorted by the trends at the lower wave numbers. As in this paper, we would prefer to reduce the preprocessing operations to a necessary minimum, the related data are not shown.

*4.2. From long to short superperiods*

To detect significant short superperiods, the scheme presented in Section 2.3 should be modified. The steps in the modified scheme are as follows. (i) Divide the genome by non-overlapping segments of length $L_{segment}$. The length $L_{segment}$ should be intermediate between the characteristic lengths corresponding to the consecutive levels of hierarchical chromosome organization. (ii) Perform DFT for DNA sequences within segments. (iii) Perform DDFT as described in Section 2.3. (iv) Pick out the spectral range in DDFT spectrum corresponding to the number of superperiods $N_{sp|segment} = 2$–10 and renormalize the range on the local mean (Eq. (40)).

(v) Average the renormalized spectral ranges over segments. If needed, the averaged spectral range should be de-trended as described in Section 2.6. After de-trending, the harmonics within the resulting spectrum should again be renormalized on the mean in the range.

At the limit of large number of segments $N_{segments} = M/L_{segment}$, the random counterparts of harmonics in averaged spectral ranges will approximately be described by Gaussian statistics with the mean 1 and the dispersion $1/N_{segments}$. Therefore, the resulting averaged spectral ranges can conveniently be analyzed in terms of deviations

$$\zeta_{II}(q_{n'}) = \left(< f_{II}(q_{n'}) >_{segments} - 1\right)/(1/N_{segments})^{1/2} \tag{45}$$

where $< f_{II}(q_{n'}) >_{segments}$ is the amplitude (40) averaged over segments (and preprocessed, if needed, as described in the item (v) above). The significant short superperiods are detected by the positive peaks in $\zeta_{II}(q_{n'})$ spectrum. The corresponding number of superperiods $N_{sp|segment}$ within segments is related to the wave number $n'$ for the peak deviation $\zeta_{II}(q_{n'})$ similarly to Eq. (25) with $N_c = [L_{segment}/2]$. The total number of related superperiods in the whole genome is given by

$$N_{sp|genome} = N_{sp|segment}(M/L_{segment}) = \left(([L_{segment}/2] - 1)/n'\right)(M/L_{segment}) \tag{46}$$

We chose $L_{segment} = 50$ kb for the *E. coli* genome ($N_{segments} = 92$).

The resulting normalized deviations (45) for DDFT spectral ranges corresponding to the short superperiods obtained for the sum (8) and for the base-pairing free energy profile at 37°C are shown in Fig. 4. The numbers of superperiods per genome were calculated by Eq. (46). DDFT spectra after averaging over segments were quadratically de-trended as described in the item (v). In the case of the sum (8), the trend is weak and can be neglected, whereas for the base-pairing free energy profile, the trend is quite visible and the related DDFT spectrum should be preliminarily preprocessed. For all highest harmonics marked in Fig. 4, there are the corresponding significant counterparts with $n = N_{sp}$ in the primary DFT spectra (Supplementary files S1 and S2).

As remarked in Section 2.3, generally, the equidistance in DFT spectrum can produce the related equidistant series in DDFT spectrum. For the chosen segment length $L_{segment} = 50$ kb, the superperiods at $N_{sp} \approx 400$ can produce two equidistant peaks in DDFT spectrum. Notably, the counterpart doublet peaks can be identified for the highest harmonic in both spectra shown in Fig. 4. The respective highest harmonics are $n' = 5326$, $N_{sp} = 435.7$, $\zeta_{II}(q_{n'}) = 4.33$ in Fig. 4A and $n' = 5217$, $N_{sp} = 444.8$, $\zeta_{II}(q_{n'}) = 4.58$ in Fig. 4B, whereas the respective nearly doubled counterparts are $n' = 10821$, $\zeta_{II}(q_{n'}) = 3.48$ and $n' = 10426$, $\zeta_{II}(q_{n'}) = 4.06$. The deviations of the wave numbers for nearly doubled counterparts from the strictly twice values lead to the uncertainty in the detected number of superperiods, $\Delta N_{sp}/N_{sp} = \Delta n'/n'$, where $n'$ should be referred to the doubled counterparts (cf. Chechetkin and Turygin, 1995a; Lobzin and Chechetkin, 2000). If the superperiods would be detected by two equidistant harmonics, the relative uncertainty in the superperiods should be, respectively, $|\Delta N_{sp}|/N_{sp} = 0.016$ and 0.001. The numbers of superperiods per genome for the doubled counterparts are, respectively $N_{sp} = 214.5$ and 222.6. In the primary DFT spectra for the sum (8) and base-pairing free energy, there are highly significant harmonics with wave numbers $n = 214, 215$ and $222, 223$. This indicates that the relevant equidistant harmonics should be attributed to the separate superperiods. We cannot exclude that both effects are superimposed. Within both interpretations, the simultaneous observation of two nearly equidistant peaks enhances strongly the statistical significance of detected superperiods.

The short superperiods corresponding to the three highest harmonics in DDFT spectra for the combinations (5)–(8) and for the base-pairing free energy profile at 37°C are summarized in Fig. 5 (the relevant details can be found in Supplementary file S2). The distributions of short superperiods for the random sequences and for the *E. coli* genome coarse-grained over intervals 50 are compared in Fig. 5B. The comparison reveals the significant bias for the superperiods with $N_{sp} \approx 300$, 500, and 800. The short superperiods are clearly clustered within the range $N_{sp} \approx 400$–600. The superperiods at $N_{sp} \approx 200$ and at $N_{sp} \approx 800$ cannot be considered as the equidistant counterparts for the superperiods with $N_{sp} \approx 400$, because they refer to the different variables. Such grouping may reflect the mechanism of period-doubling typical of modifications in genomic DNA sequences (Chechetkin and Turygin, 1995a; Lobzin and Chechetkin, 2000). Genetically, the period-doubling is related to specific modifications in DNA sequences concordant with the pitch of a seed period. These modifications may be related to regulation mechanisms and/or to unification/fragmentation of particular domains/units. The coordinated modifications of genomic DNA sequences produce an intricate network of interrelated periodicities rather than a set of singular periodic patterns.

During starvation and stress, the supercompaction of the *E. coli* nucleoid occurs that protects DNA under unfavorable environmental conditions (Meyer and Grainger, 2013). The supercompaction can be naturally achieved by the smaller structural units associated with the increase in the number of short superperiods. Therefore, the superperiods at $N_{sp} \approx 500$–600 could reflect stress-induced DNA supercompaction in *E. coli*. If the ratio about ten between the number of units at the third and second levels of the nucleoid folding would be retained after supercompaction, the number of the medium superperiods should be increased in the same proportion as the number of the short superperiods. Thus, the medium superperiods with $N_{sp} \approx 60$–70 (Fig. 3B) may also be attributed to the supercompaction.

*4.3. Temperature effects and supercoiling*

Additionally, we calculated DDFT spectra for the base-pairing free energy profiles at the temperatures 10, 20, 37, and 50°C. The main superperiods turned out to be persistent in this temperature range; however, small variations in the corresponding amplitudes as well as the permutations in the ranking of amplitudes may indicate the related variations in the nucleoid conformations with temperature (Supplementary files S1 and S2). As the genetic consequences of small variations may be significant, we studied separately the variations in the counterpart harmonics in DFT spectra for the base-pairing free energy profiles at the different temperatures $t_1$ and $t_2$,

$$\Delta f_{t_1|t_2}(q_n) = f_{t_1}(q_n) - f_{t_2}(q_n) \qquad (47)$$

The hidden regularities in such variations were assessed with DDFT for the variables (47) as described in Section 2.3. The periodicities related to variations (47) will be termed modulational superperiods. These modulations are imposed on an initial free energy profile under temperature shift.

The free energy profile at 37°C was chosen as a reference. Then, the modulational superperiods corresponding to the shifts from reference temperature to 10, 20, and 50°C were studied with DDFT for the variables (47). The results for the modulational superperiods are summarized in Fig. 6 and Supplementary file S2. As the trends in all relevant DDFT spectra were absent, de-trending was not applied. The variations (47) under shifts to lower and higher values from the reference temperature were significantly correlated. The corresponding Pearson correlation coefficients are $k\left(\Delta f_{50°C|37°C}(q_n) | \Delta f_{37°C|20°C}(q_n)\right) = 0.691$; $k\left(\Delta f_{50°C|37°C}(q_n) | \Delta f_{37°C|10°C}(q_n)\right) = 0.766$; and $k\left(\Delta f_{37°C|20°C}(q_n) | \Delta f_{37°C|10°C}(q_n)\right) = 0.992$ (Pr < 0.001).

The relationship between the significant modulational superperiods and the quasi-periodic expression patterns (Table 1) may be closer in comparison with superperiods. The genes in Bacteria are commonly actively transcribed under negative supercoiling in bacterial chromosome (Hatfield and Benham, 2002; Deng et al., 2005; Travers and Muskhelishvili, 2005, 2013; and references therein). Physically, both increase of temperature and negative supercoiling result in weakening of base-pairing in dsDNA. In this sense, the increase of temperature can mimic the supercoiling effects, whereas the inhomogeneous distribution of temperature over the genome can model the inhomogeneous supercoiling. The decrease of temperature can represent positive supercoiling typical of hyperthermophilic Archaea (Musgrave et al., 1991; Grayling et al., 1996; Valenti et al., 2011). Within such approach, the temperature should be treated formally as a fitting parameter simulating level of supercoiling. The analogue temperature for the highest level of negative supercoiling should be not far from the local melting temperature of dsDNA. Thus, the superperiodic variations in base-pairing free energy profile under the temperature shift can model the counterpart variations under supercoiling.

The correspondence between the most significant short modulational superperiod under the temperature shift to 50°C with $N_{sp} = 286.5$ and the short-range expression patterns appeared to be better than that for the counterpart free energy superperiod, whereas the correspondence between the long and medium modulational superperiods and relevant expression patterns retained approximately the same in comparison with the counterpart superperiods. The short modulational superperiods with $N_{sp} \approx 600$ were present among the most significant superperiods for all temperature shifts and may be related to the stress-induced supercompaction of the *E. coli* nucleoid. Taking into account the parallels between increase of temperature and negative supercoiling, we also calculated the significant modulational superperiods under the shifts to very high temperatures 60, 70, and 80°C. Such temperatures are beyond the functional temperature range for *E. coli* and should be treated formally as a model of supercoiling. The relevant data are presented in Supplementary file S2 and reveal the higher impact of the shorter length scales for the significant modulational superperiods at the increase of temperature/supercoiling. Note the absence of trends in the related DDFT spectral ranges for the variations (47) at the shifts to 60, 70, and 80°C as well. The most significant long and medium modulational superperiods at the shift to 60°C with $N_{sp} = 2.2; 5.9; 3.2$ and $N_{sp} = 38.5; 40.7; 35.2$ appeared to be very close to their counterparts in Table 1, whereas the short modulational superperiods with $N_{sp} = 286.5; 598.1; 311.8$ were close to the short-range expression patterns, except superperiods with $N_{sp} = 598.1$ being related to the nucleoid supercompaction (see also spectra in Supplementary file S3). Such correspondence indicates that the mean analogue temperature for the negatively supercoiled *E. coli* chromosome is about 60°C in rapidly growing bacteria.

*4.4. Correlations in the E. coli genome*

The characteristic length scales in the genome organization are reciprocally mapped by Fourier structure factors and correlation functions. In this aspect, the structure factors and the correlation functions are complementary to each other (Lobzin and Chechetkin, 2000). Nevertheless, the information stored in both structure factors and correlation functions is the same, because they are related by the Wiener-Khinchin relationship (11). In this section, we present some results on the correlations in genomic DNA sequence of *E. coli*.

Fig. 7 shows the dependences on separation distance $m_0$ for the sum of normalized deviations $Q_4(m_0)$ (Eq. (18)) and for the normalized deviations for base-pairing free energy $\kappa_{FE}(m_0) = (K_{FE}(m_0) - \overline{K}_{FE}) / <\Delta K_{FE}^2>_{random}^{1/2}$ (Eqs. (29)–(32)). The difference between a current value of correlation function and its mean value can be presented as

$$K_{\alpha\alpha}(m_0) - \overline{K}_{\alpha\alpha} \approx \frac{1}{M} \sum_{m=1}^{M} (\rho_{m,\alpha}^c - \overline{\rho}_\alpha)(\rho_{m+m_0,\alpha}^c - \overline{\rho}_\alpha); \quad \overline{\rho}_\alpha = N_\alpha / M \qquad (48)$$

The positive correlations (or correlations) correspond to the variations $\rho_{m,\alpha}^c - \overline{\rho}_\alpha$ and $\rho_{m+m_0,\alpha}^c - \overline{\rho}_\alpha$ of the same sign, whereas the negative correlations (or anticorrelations) correspond to the variations $\rho_{m,\alpha}^c - \overline{\rho}_\alpha$ and $\rho_{m+m_0,\alpha}^c - \overline{\rho}_\alpha$ of the different signs. Fig. 7A reveals a slight but distinct genome-scale trend from correlations to anticorrelations for $Q_4(m_0)$ at the increase of $m_0$. The mean value of correlation function for nucleotides depends only on the nucleotide composition and does not depend on the distribution of nucleotides over the genome (Eq. (13)). As the mean value is fixed, the significant correlations at the shorter separation distances $m_0$ produce the anticorrelations at the longer distances and vice versa. Or in other words, as the total number of nucleotides of particular type is fixed in the genome, if the pairs of separated nucleotides are encountered more frequently at the shorter separation distances, the pairs of nucleotides will be encountered less frequently at the longer distances. The trend in Fig. 7A cannot be attributed to the strong correlations at $m_0 < 5$ kb (Fig. 7B), because in this case the same trend should be observed in Fig. 7C as well. Thus, the trend in Fig. 7A reflects the genome-scale correlations between nucleotides. Though the characteristic length is seemingly absent in the general trend, there is a set of characteristic lengths related to the outbursts in correlations/anticorrelations.

Up to $m_0$ about 2.5 kb, the correlations dominate over anticorrelations for both $Q_4(m_0)$ and $\kappa_{FE}(m_0)$ (Figs. 7B and 7D). The characteristic lengths for the drop of correlations for $Q_4(m_0)$ and $\kappa_{FE}(m_0)$ assessed by the intercepts with the 5% significance threshold for the Gaussian extreme value statistics are, respectively, about 2.5 and 5.5 kb. Note that the correlation length for the base-pairing free energy profile turns out to be distinctly longer than that for the deviations in nucleotide composition. Such drop of correlations is partially related to the variations in frames for periodic patterns with the three-base period $p = 3$ typical of protein-coding regions (Chechetkin and Turygin, 1995b; Herzel and Große, 1997). The three-periodic patterns are clearly seen in the insert to Fig. 7D (see also the clusters of high peaks in the vicinity of $p \approx 3$ for the different DFT spectra in Supplementary file S1). The characteristic lengths about 1 kb can be referred to (i) length of genes; and (ii) length of Okazaki fragments at the synthesis of lagging strand during replication (Lewin, 2000). The total propagation distance for transcription-induced supercoiling in the upstream and downstream vicinity of a transcribed region in *E. coli* comprises $\approx 0.8 + 4 = 4.8$ kb (Wu et al., 1988; Rahmouni and Wells, 1992; Moulin et al., 2005) and appears to be quite close to the correlation length for $\kappa_{FE}(m_0)$. The finite-range correlations with the correlation length $l_{corr}$ like that shown in Figs. 7B and 7D induce the trends in Fourier spectra at the low wave numbers up to $n \cong M / l_{corr}$ (Lobzin and Chechetkin, 2000). The replacement of the normalized deviations by the respective $\chi^2$-criteria does not affect the main dependences shown in Fig. 7.

As the free energy profiles are calculated with the energy for tiling dinucleotides (Section 2.4), this leads to the sharp correlations at $m_0 = 1$ (insert to Fig. 7D) present even in the random sequences. Beyond separation distances $m_0 > 1$, the related correlations in the random sequences would be absent. Let the pairing between DNA strands be significantly less strong within the short stretches of characteristic length $l_{ch}$ relative to that in the surrounding vicinity. This situation is typical of the sites for transcription initiation (Rangannan and Bansal, 2010) and is also favorable for binding of DNA strand with RNA primers at the lagging strand synthesis during the replication (Robinson and van Oijen, 2013). Then, the anticorrelations should be observed for the range of separation distances $m_0 > l_{ch}/3$, if the contribution of correlations from the surrounding vicinity is relatively small. Such anticorrelations are seen in the insert to Fig. 7D in the range around $m_0 \approx 5$. For the *E. coli* genome, the three-base correlations at $m_0 > l_{ch}$ will rapidly overcome the contribution of such anticorrelations.

Returning to the search for hidden large-scale regularities in genomic DNA sequences with the use of correlation functions, it would be nearly impossible to discern them in the variations like that shown in Figs. 7A and 7C. At least, such regularities can be displayed only after sophisticated preprocessing of the primary data. Unlike Fourier transform, the separation of the particular components in the superposition $a_1\cos(2\pi m/(M/2)+\varphi_1) + a_2\cos(2\pi m/(M/3)+\varphi_2) + a_3\cos(2\pi m/(M/6)+\varphi_3) + noise$ with different amplitudes and phases would be uneasy using the correlation function. This shows the efficiency of DDFT in search for large-scale regularities in comparison with the other methods.

*4.5. Example: the writhed structure*

The direct consequences of DDFT analysis for the large-scale chromosome architecture may be illustrated with a model presentation of the writhed structure corresponding to the highest level of the *E. coli* nucleoid folding (Umbarger et al., 2011; Dorman, 2013). To stress the salient features, the complicated hierarchical structure of the *E. coli* chromosome will be schematically replaced by a thick elastic rod. Let one end of the rod be double-turned around the axis and connected to form a ring. The resulting structure is shown in Fig. 8. The distribution of double turn between writhing and twisting will be implied approximately equal. Let $2\pi$ twisting be approximately homogeneously distributed between the two $\pi$-turn regions marked in Fig. 8. Within such regions, the cross-section rotates on $\pi$.

The semi-elliptical segments in Fig. 8 form repeating structural units for the writhed structure and can approximately be referred to the genetic macrodomains (Valens et al., 2004; Dorman, 2013). The long superperiods with $N_{sp} \approx 6$ can be juxtaposed to these repeating structural units. The alternating curvature in the semi-elliptical segments of the writhed structure can be produced by alternating A- or T-tracts in the apex regions of semi-ellipses. The similar effect can be achieved by architectural proteins bound to specific DNA motifs or by the combination of both mechanisms (the latter situation is most probable). The distribution of A- or T-tracts along the writhed structure can be juxtaposed to the long superperiods with $N_{sp} \approx 3$ which should be seen in DDFT spectra for $f_{AA}(q_n)$, $f_{TT}(q_n)$, and $f_{AA}(q_n) + f_{TT}(q_n)$ (though the corresponding superperiods for $f_{AA}(q_n)$ are absent in Supplementary file S2, they are present among top five ranked harmonics). This explains the simultaneous co-existence of superperiods with $N_{sp} \approx 3$ and 6. They reflect the different aspects of folding for the writhed structure and polar difference between the upper and lower domains. Generally, the co-existence of superperiods can be related to the superimposed binding profiles for the different families of architectural proteins. In particular, the binding of proteins to specific DNA motifs in the vicinity of loci *oriC* and *dif* (or protein-DNA binding specific to macrodomains Ori and Ter) may be related to the superperiods with $N_{sp} \approx 2$. As the sites of protein binding are coordinated with the quasi-regular chromosome architecture, the superperiods appear often to be commensurate. The relationship between such commensurability and various genetic mechanisms needs additional study.

*4.6. Other genomes*

The well-studied *E. coli* chromosome was chosen as a main test for the general scheme of DDFT analysis. Two other examples briefly presented in this section serve merely for illustration of further abilities of DDFT. The first example concerns the superperiods in the genome of bacteriophage PHIX174 (RefSeq assembly accession GCF_000819615.1; $M = 5,386$; $N_A = $

1,291; $N_G = 1,254$; $N_T = 1,684$; $N_C = 1,157$). The single-stranded circular DNA of PHIX174 encoding 11 genes is confined within dodecahedron viral capsid having the icosahedral symmetry (Fig. 9A, on the left). Remind that the symmetry groups for dodecahedron and icosahedron are the same despite the difference in the figures. As the capsid affects the scaffolding of the genome (Dokland et al., 1999; Hafenstein and Fane, 2002), the icosahedral symmetry can be displayed in the underlying genome sequence. The axes of the fifth, third, and second order are inherent to the icosahedral symmetry. Therefore, the corresponding superperiods are expected to be observed in the related DDFT spectra for DNA sequence. Indeed, the high harmonics in DDFT spectrum for $S_4(q_n)$ (Eq. (8)) indicate such superperiods (Fig. 9A, on the right). Note that the doublet peaks at $N_{sp} = 5.4$ and 2.6 may be approximately assigned to the superperiods with $N_{sp} \approx 5.4$. On the other hand, the superperiods with $N_{sp} \approx 5.4$ may approximately be attributed to the period-doubling of the superperiods with $N_{sp} \approx 11.3$ (note also the correspondence between the superperiods with $N_{sp} \approx 11.3$ and the total number of genes). The hidden equidistance in the primary DFT spectra related to the symmetry of the fifth order can also be identified by the sums of harmonics (Chechetkin and Turygin, 1995a; Lobzin and Chechetkin, 2000).

The capsids for the majority of viruses form either helical or icosahedral structure. This leads to the correspondence between superperiods in the genomic DNA and either a helix pitch for helical capsid structure or the order of symmetry for icosahedral structure. The rule that the scaffolding of the viral genome should be coordinated with the symmetry of capsid allows to assess the related number of superperiods in the underlying genomic sequences and vice versa. As the details of the genome scaffolding within capsid are often difficult to resolve via X-ray analysis or electron microscope imaging, the related superperiods detected by DDFT may appear to be helpful for the preliminary assessment of genome scaffolding. The relationship between such large-scale viral genome organization and various genetic mechanisms needs separate investigation.

The second example concerns *Caulobacter crescentus* CB15 chromosome (GenBank assembly accession GCA_000006905.1; $M = 4,016,947$; $N_A = 661,121$; $N_G = 1,347,429$; $N_T = 655,876$; $N_C = 1,352,521$). Using 5C-based technique, Umbarger et al. (2011) resolved large-scale folding of this bacterium chromosome. The irregular writhed structure can be attributed to the double-twisted type shown in Fig. 8. Similarly to the *E. coli* genome, the DDFT spectrum for *C. crescentus* reveals the characteristic high harmonics with superperiods about $N_{sp} \approx 6$, 3, and 2 as well (Fig. 9B). The corresponding peaks for *C. crescentus* are less pronounced than their counterparts for *E. coli*; this can be related to the strong AT/GC bias in the *C. crescentus* genome. The detailed comparative study of different genomes is planned at the next stage of investigations.

## 5. Discussion

Our study proves that DDFT is able to detect large-scale quasi-periodic regularities in genomic DNA sequences. The detected superperiods in the *E. coli* genome revealed good correspondence with the domains/units found experimentally. The clear clustering of superperiods correlates with the hierarchical folding of the *E. coli* nucleoid. The co-existence of different superperiods at given level of folding indicates the dynamic variations in the nucleoid conformations at the different phases of cell cycle and/or at the response to the change in the environmental conditions. Our results indicate also the superimposed macrodomain organization in the *E. coli* genome. The detailed experiments and modeling like that of performed by Umbarger et al. (2011) could further elucidate the relationships between the superperiods and the dynamic domains/units in bacterial chromosomes. At the next step, we intend to apply DDFT methods to the eukaryotic genomes.

Currently, wavelet transform is the main tool for the study of large-scale genome organization. Allen et al. (2006) used Morlet wavelet transform to search for large-scale genome regularities in the *E. coli* and other bacterial genomes. They detected 600–650-kb quasi-periodic patterns in a scalogram for the *E. coli* genome. These patterns correspond to 7.1–7.7 superperiods, whereas DDFT detects 6.3 superperiods (Fig. 2). The latter observation appears to be in the better agreement with the number of large-scale units in the writhed structure (Fig. 8), whereas the former one is in the better correspondence with the long-range expression patterns (Table 1). Allen et al. (2006) observed also periodic patterns at length scales 80–100 kb corresponding to 46.4–58.0 medium superperiods, which are close to superperiods detected by DDFT and shown in Fig. 2C. The wavelet data for short superperiods were not reported and cannot be compared with their DDFT counterparts. The detection of the superperiods via DDFT is much sharper in comparison with that for wavelet transform. The locality of analysis via wavelet transform cannot be considered as the main advantage in comparison with DDFT, because DDFT can be performed within the windows moving over the genome as well. Unlike DDFT spectra, analytical criteria for the assessment of the statistical significance of the salient features in the wavelet scalograms are absent. The significance can be assessed only by the comparison with randomized sequences. The assessment should be performed by the extreme value statistics and needs at least $10^3$–$10^4$ random realizations. The use of the multiple random realizations makes the computational time much longer in comparison with the algorithms based on the analytical criteria (Kravatsky et al., 2015). The computational complexity of Fourier and wavelet transforms is comparable. Note that the fast Fourier transform imposing limitations on the sequence length should be used with care and was not applied in this paper.

The most pronounced long and short superperiods in the *E. coli* genome turned out to be related to the combination $f_{AA}(q_n) + f_{CC}(q_n)$ (Supplementary file S2). The second ranked long superperiods corresponded to $f_{AA}(q_n) + f_{TT}(q_n)$. The latter combination is known to be related to dsDNA curvature and transcription regulation (Pérez-Martín and De Lorenzo, 1997; Trifonov, 1998; Jáuregui et al., 2003; Klaiman et al., 2009; Kravatskaya et al., 2011, 2013; and references therein). The curvature effects are commonly associated with the helical AT-periodicity near $p =$ 10.5. If the stretches with pronounced helical AT-periodicity were coordinately positioned over the genome, such positioning could be related to the large-scale genome organization. The similar effects for the periodicity $p = 3$ have been discussed by Chechetkin et al. (1994). As the periodicity $p = 3$ is commonly strict within particular gene for the nucleotides of at least one type, the distribution of stretches with differently phased periodicity $p = 3$ reflects the positioning of genes over the genome. Therefore, though both periodicities $p = 3$ and 10.5 are short-ranged, the clusters of high peaks around these periods in DFT spectrum can affect the detection of superperiods (and actually do in some cases, as shown by our computations with variable cut-off $N_c$). Although the peaks in DDFT spectrum for the variable $S_4(q_n)$ (Eq. (8)) are not among the highest for all set (5)–(8), the numbers of superperiods detected by $S_4(q_n)$ appear to be in the best correspondence with domains/units listed in Table 1. Yet, using together the set of all variables (5)–(8) and the base-pairing free energy profile provides more complete information on the large-scale genome organization than that inferred from the spectrum for particular variable.

DDFT analysis of base-pairing free energy profiles (together with profiles for bending rigidity of dsDNA) allows to assess variations in the chromosome conformations at different temperatures. The increase of temperature can also model the effects of negative supercoiling. In addition to the combinations of nucleotides and base-pairing free energy, it would be useful to perform DDFT analysis for the profile of dsDNA curvature and for the profiles of binding nucleoid-associated proteins to dsDNA to elucidate large-scale nucleoid organization. The combinations of binding profiles for different proteins can be treated similarly to the sums (6)–(8). As genomic DNA sequences remain the major source in the study of the molecular evolution of the different organisms, DDFT approach may provide the valuable information on the evolution of the large-scale genome organization. DDFT analysis of large-scale genome variations between two aligned genomic sequences can be performed similarly to the scheme presented in Section 4.3.

Let the total length of the genome be $M$. The statistical methods can detect regularities in genomic DNA sequences if there is a regular bias in the types of nucleotides (or in specificity of nucleotides) occupying more than $2\sqrt{M}$ sites coordinately positioned over the genome. In particular, for the *E. coli* genome the threshold $2\sqrt{M}$ is about 4,100 bp. Taking into account the scattering both in the specificity of nucleotides and in their positioning, the actual threshold will be in part higher and $2\sqrt{M}$ should be considered as the lower limit. If the proteins recognize specifically the DNA motifs of about 10–20 bp in length, the lower limit on the number of proteins bound specifically to the coordinated positions of the genome should be 200–400 to satisfy the detection threshold $2\sqrt{M}$. This estimate concerns both the architectural proteins and the macrodomain markers in the *E. coli* genome.

Many eukaryotic genomic DNA sequences are assembled with the lacunae. The estimate like $2\sqrt{M}$ concerns also the total length of the lacunae. If the length of the lacunae exceeds $2\sqrt{M}$, they may affect the detection of superperiods by DDFT. For the human genome $M \approx 3 \times 10^9$ bp, and the total length of the lacunae should not exceed 110 kb. The actual total length of the lacunae for the human genome exceeds this threshold, that should be taken into account at the bioinformatic analysis.

The thresholds based on the extreme value statistics provide reasonable criteria for the significance assessment of the large-scale periodicities with DDFT. As the ratio $f/f_{thr}$ is insensitive to the number of harmonics in the chosen spectral range, the normalization of Fourier amplitudes on $f_{thr}$ may be useful when comparing counterpart features for the large-scale periodicities in genomes of different length. If the amplitudes in the chosen spectral range do not exceed the extreme value threshold, the choice of the highest peaks provides approximate assessment of the large-scale periodicities. Nevertheless, even in this case, the peaks should not be far from the thresholds. The thresholds can be lowered for the reproducible features and/or at the additional correlations like the simultaneous observation of nearly equidistant peaks in the same spectrum. The persistence of superperiods for the nucleotides of different types as well as for the combinations of different nucleotides enhances the statistical significance of detected large-scale genome regularities. The collected criteria of large-scale periodicities comprise (i) high significant peaks (preferably isolated, i.e. considerably higher than nearby peaks) in the primary DFT spectrum; (ii) corresponding high peaks in the counterpart DDFT spectrum; (iii) reproducibility of peaks for the nucleotides of different types and for their different combinations; and (iv) simultaneous detection of commensurate peaks (with the ratio of wave numbers close to an integer) in the same or different spectra.

It is important that the chromosome architecture affects the gene expression. The periodic expression patterns (Jeong et al., 2004) and their correspondence with the structure of the *E. coli* nucleoid (Dorman, 2013) show that the expression of genes obeys approximately the proximity rule: if the genes are spatially close (either along the DNA strand or in three dimensions), the levels of their expression are similar to each other. The reasons of the proximity rule may be kinetic. After switching off the transcription of a particular gene, the transport of transcription factors and molecular subunits for the assembly of RNA polymerase to switch on the transcription of a nearby gene would be much more optimal than the molecular transport from remote cellular regions. The control of expression depends on coordinated molecular interactions, among them, the specific recognition of sequence elements within promoter DNA by sigma subunit of RNA polymerase (Browning and Busby, 2004; Hook-Barnard and Hinton, 2007). The sigma subunits of seven different types are known for *E. coli*. Relative to transcription level, the promoters are often divided into strong, medium, and weak categories. Evidently, the promoters of similar strength would be preferable for such sharing mechanism. The spatial closeness of genes has also the advantages for their co-regulation (Dorman, 2013). Wright et al. (2007) screened more than 100 bacterial genomes to detect the pairs of genes that have a tendency to co-occur and a tendency to be located close together in many genomes. Analyzing the distribution of such pairs in *E. coli*, they found that genes in a pair tend to be separated by integral multiples of 117 kb along the genome. This proves that the proximity rule determines also the evolutionary selection for co-location of genes along the genome. Reciprocally, the observation of such patterns indicates the helix-like topology with pitch of 117 kb for nucleoid folding.

Though in this paper we were primarily interested in the relationships between large-scale quasi-periodic patterns in genomic DNA sequences and large-scale chromosome folding, the DDFT method developed for the analysis of such problem is more universal. Formally, the applicability of DDFT needs a large enough array of sampling points taken frequently in comparison with noisy quasi-periodic variations. Then, this array is used for the statistical assessment of large-scale hidden periodicity. Such problem formulations can be encountered in population dynamics, physics, engineering, and economics. We envisage the applications of DDFT in these fields as well.

**Acknowledgements**

The authors are grateful to Y.V. Kravatsky for the help at the early stages of investigation and figure preparation and to E.V. Lobzina for the help in the preparation of figures.

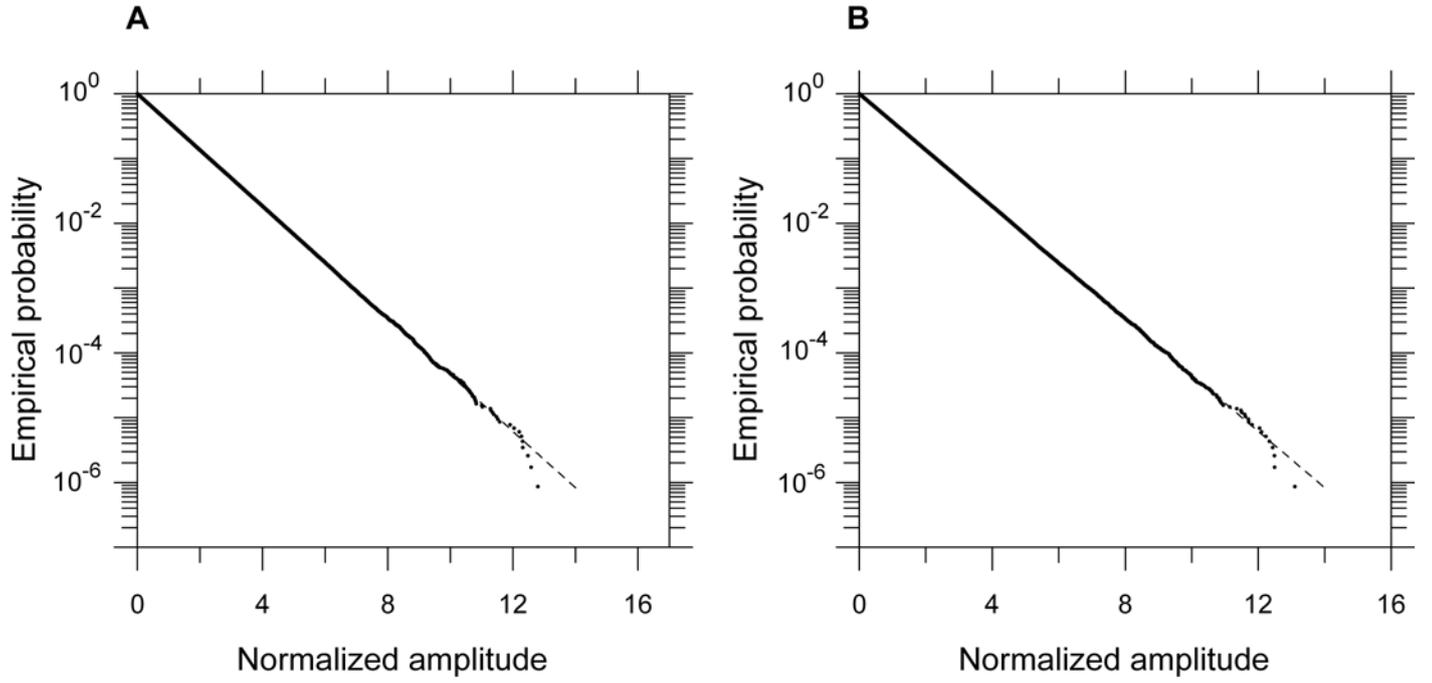

**Fig. 1.** Empirical probability distribution $P(f_{II}) = N(f'_{II} > f_{II})/N'$, where $N(f'_{II} > f_{II})$ is the number of amplitudes in DDFT spectrum exceeding current threshold $f_{II}$ and $N'$ is the total number of amplitudes in DDFT spectrum. (**A**) The empirical probability for DDFT of structure factors $f_{AA}(q_n)$ (Eq. (5)); and (**B**) The empirical probability for DDFT of $S_4(q_n)$ (Eq. (8)). DDFT spectra were obtained for the random sequence of length $M = 4,641,652$ with nucleotide composition $N_A = 1,142,742$; $N_G = 1,177,437$; $N_T = 1,141,382$; and $N_C = 1,180,091$ as described in Section 2.3. The length and nucleotide composition for the random sequence coincide with that of *E. coli* K-12 genome. The broken line corresponds to the Rayleigh probability (Eq. (36)).

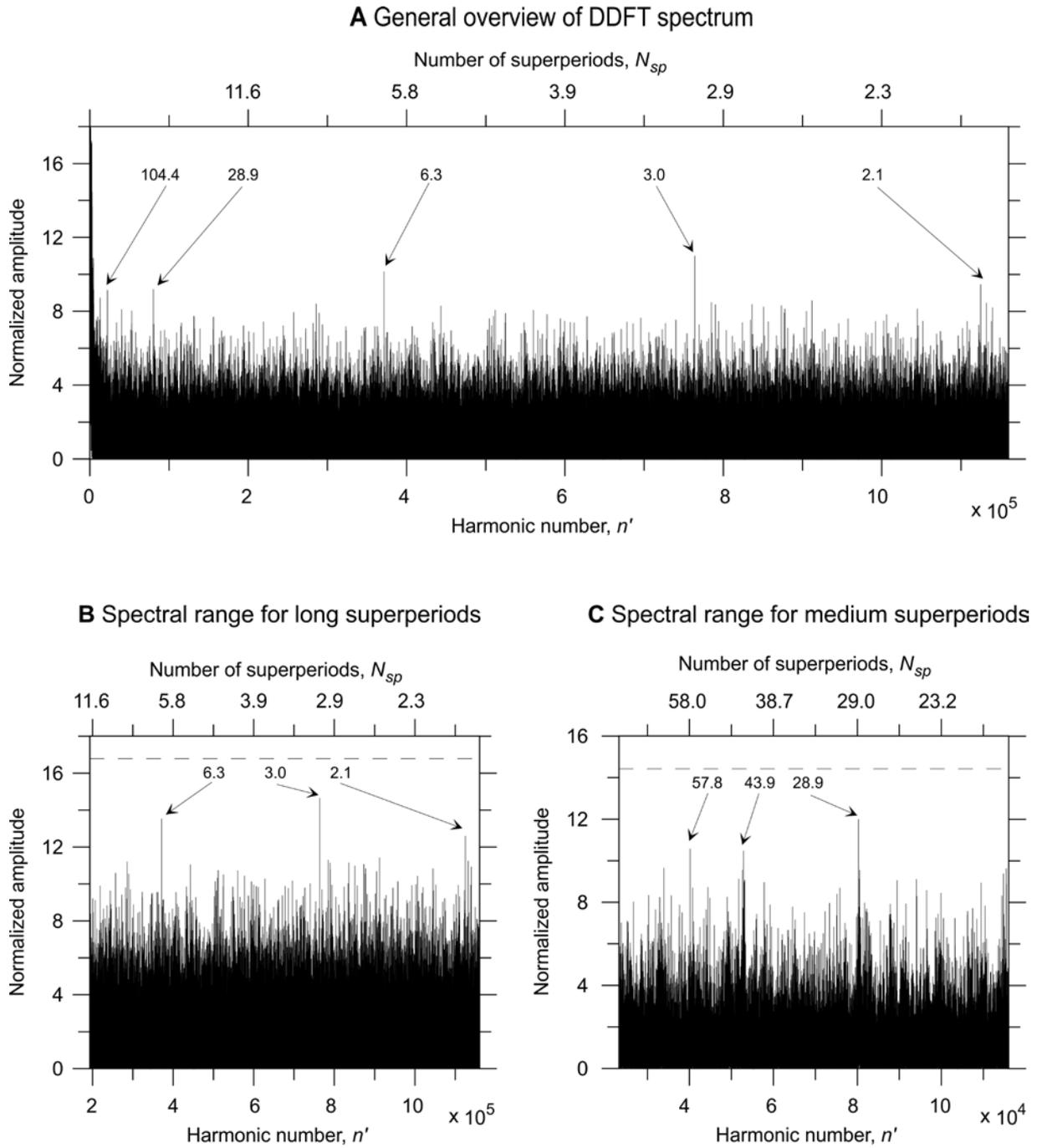

**Fig. 2.** DDFT spectrum (**A**) and the spectral ranges for the long (**B**) and medium (**C**) superperiods corresponding to the sum $S_4(q_n)$ (Eq. (8)). The horizontal broken lines refer to 5% significance thresholds for the Rayleigh extreme value statistics in the chosen spectral ranges (Section 2.5). The three highest harmonics in DDFT spectral ranges are indicated by arrows and the numbers of corresponding superperiods are shown explicitly. The long superperiods related to the three highest harmonics are consistent with the large-scale genome macrodomains and the largest structural units participating in the nucleoid folding for *E. coli*, whereas the medium superperiods with $N_{sp} = 43.9$ are rather close to the periodicity 117 kb (Table 1).

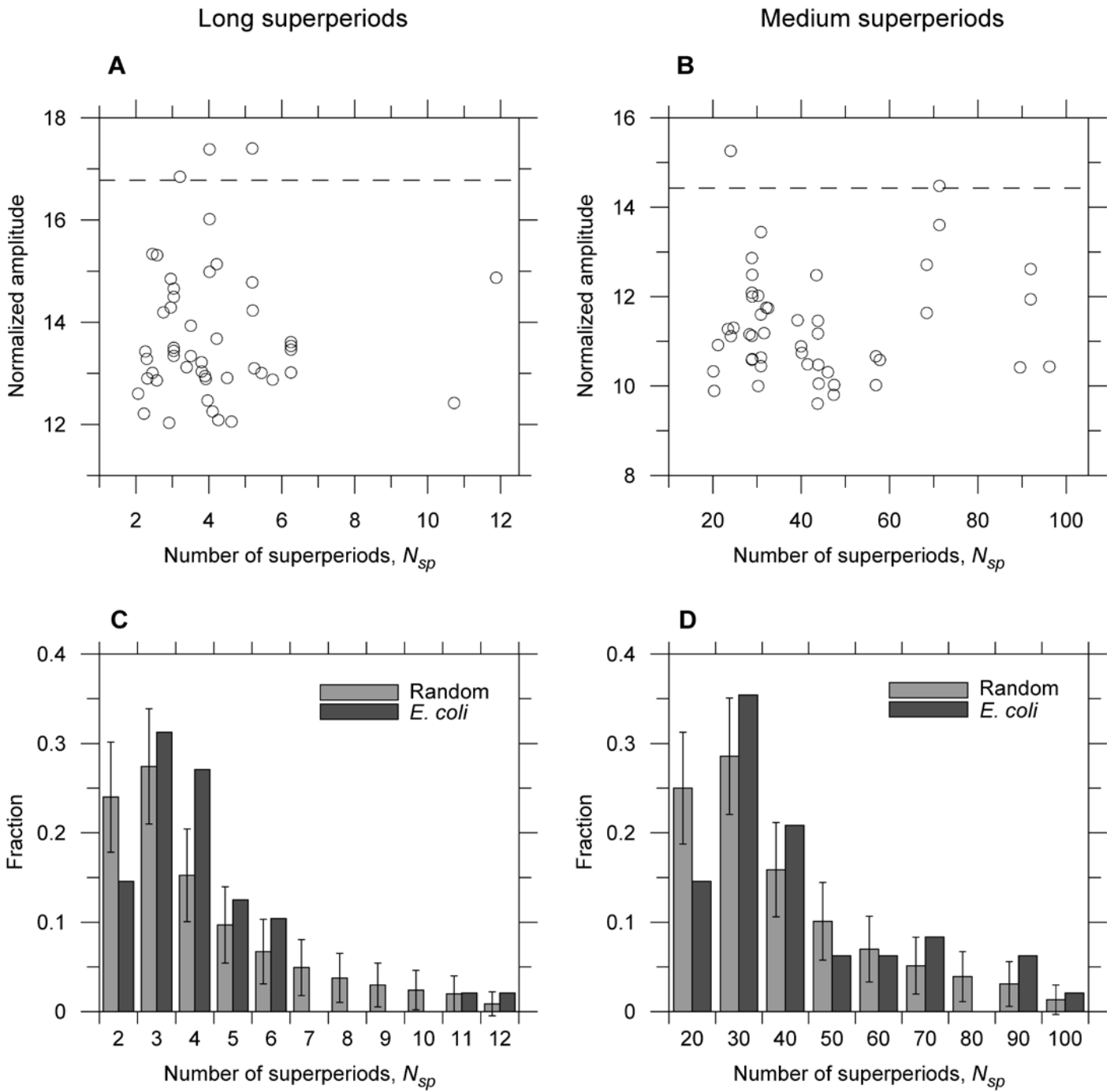

**Fig. 3.** The numbers of long (**A**) and medium (**B**) superperiods related to the three highest harmonics in the respective DDFT spectral ranges for the variables (5)–(8) and for the base-pairing free energy profile at 37°C. The horizontal broken lines refer to 5% significance thresholds for the Rayleigh extreme value statistics in the chosen spectral ranges (Section 2.5). The coarse-grained distributions of long and medium superperiods for the random sequences and the *E. coli* genome are compared in panels **C** and **D**, respectively. The data in this figure refer to domains/units at the two largest levels of folding for the *E. coli* nucleoid (see the main text and Table 1).

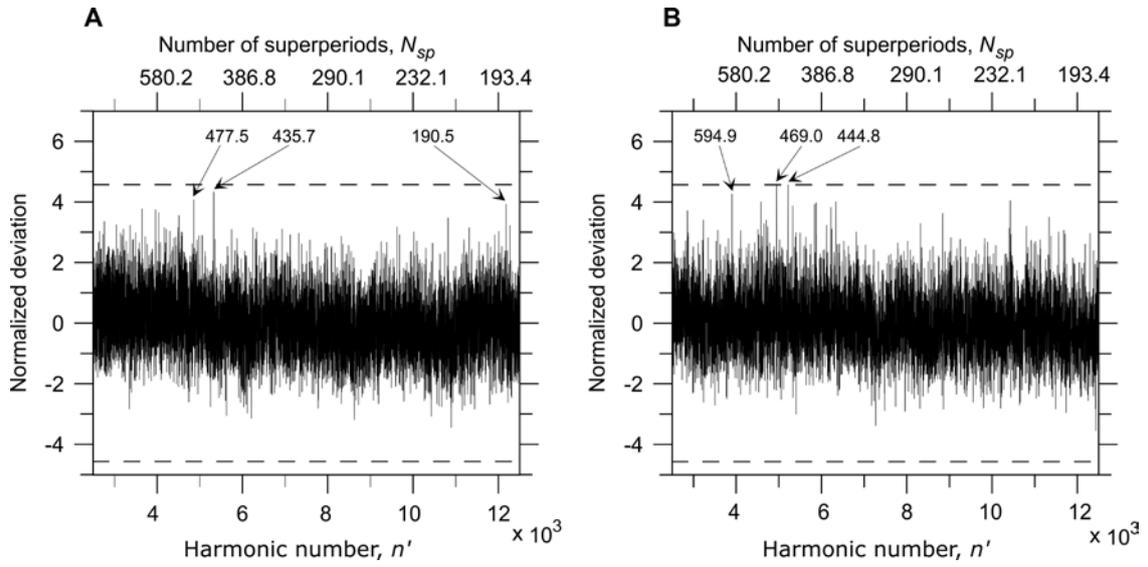

**Fig. 4.** DDFT spectral ranges for the short superperiods corresponding to the sum $S_4(q_n)$ (Eq. (8)) (**A**) and to the base-pairing free energy profile at 37°C (**B**). DDFT spectra were processed as described in Section 4.2, quadratically de-trended after averaging over segments (Eq. (44)), and transformed to the normalized deviations (Eq. (45)). The horizontal broken lines refer to 5% significance threshold for the Gaussian extreme value statistics in the chosen spectral range (Section 2.5). The three highest harmonics in DDFT spectra are indicated by arrows and the numbers of corresponding superperiods per genome (Eq. (46)) are shown explicitly. The superperiods with $N_{sp} \approx 400$–$500$ correspond to the microdomains at the third level of folding for the *E. coli* nucleoid (Table 1).

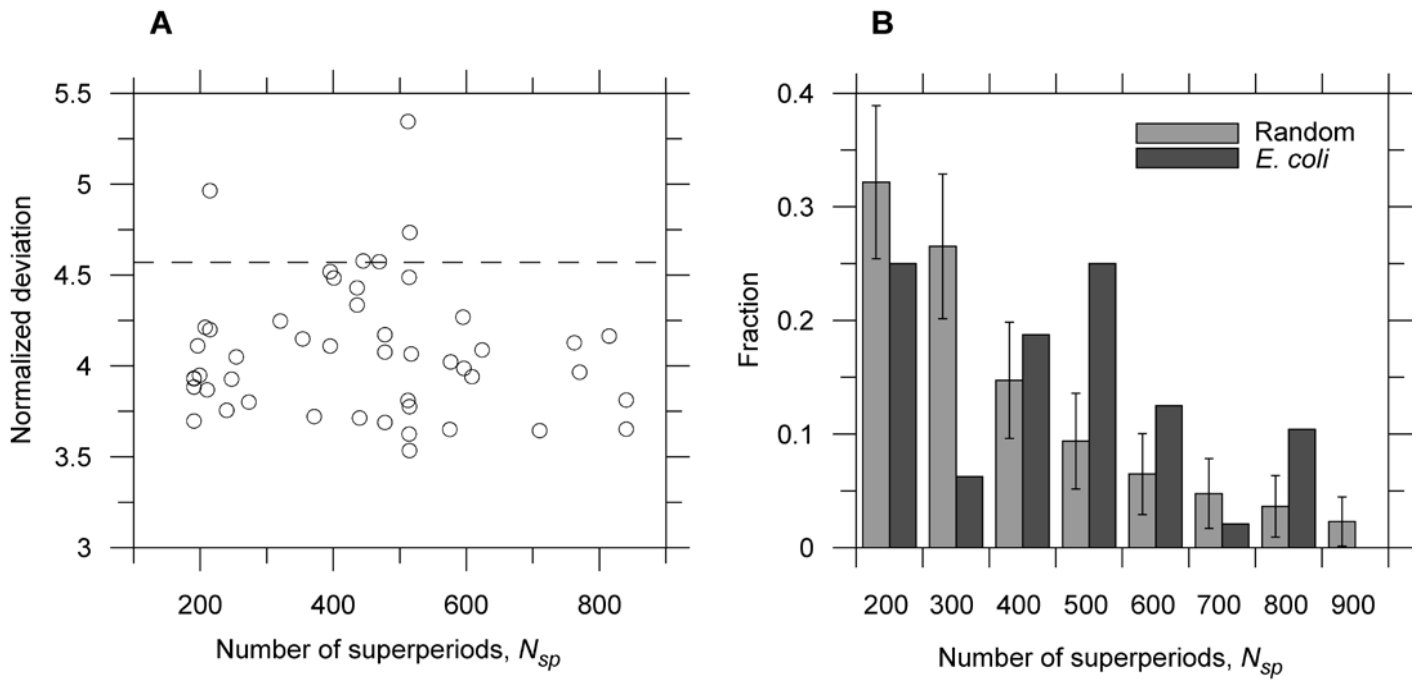

**Fig. 5.** The numbers of short superperiods corresponding to the three highest harmonics in DDFT spectra for the variables (5)–(8) and for the base-pairing free energy profile at 37°C (**A**). The DDFT spectra were processed as described in Section 4.2, quadratically de-trended after averaging over segments (Eq. (44)), and transformed to the normalized deviations (Eq. (45)). The horizontal broken line refers to 5% significance threshold for the Gaussian extreme value statistics in the chosen spectral range (Section 2.5). The coarse-grained distributions of the short superperiods for the random sequences and the *E. coli* genome are compared in panel **B**.

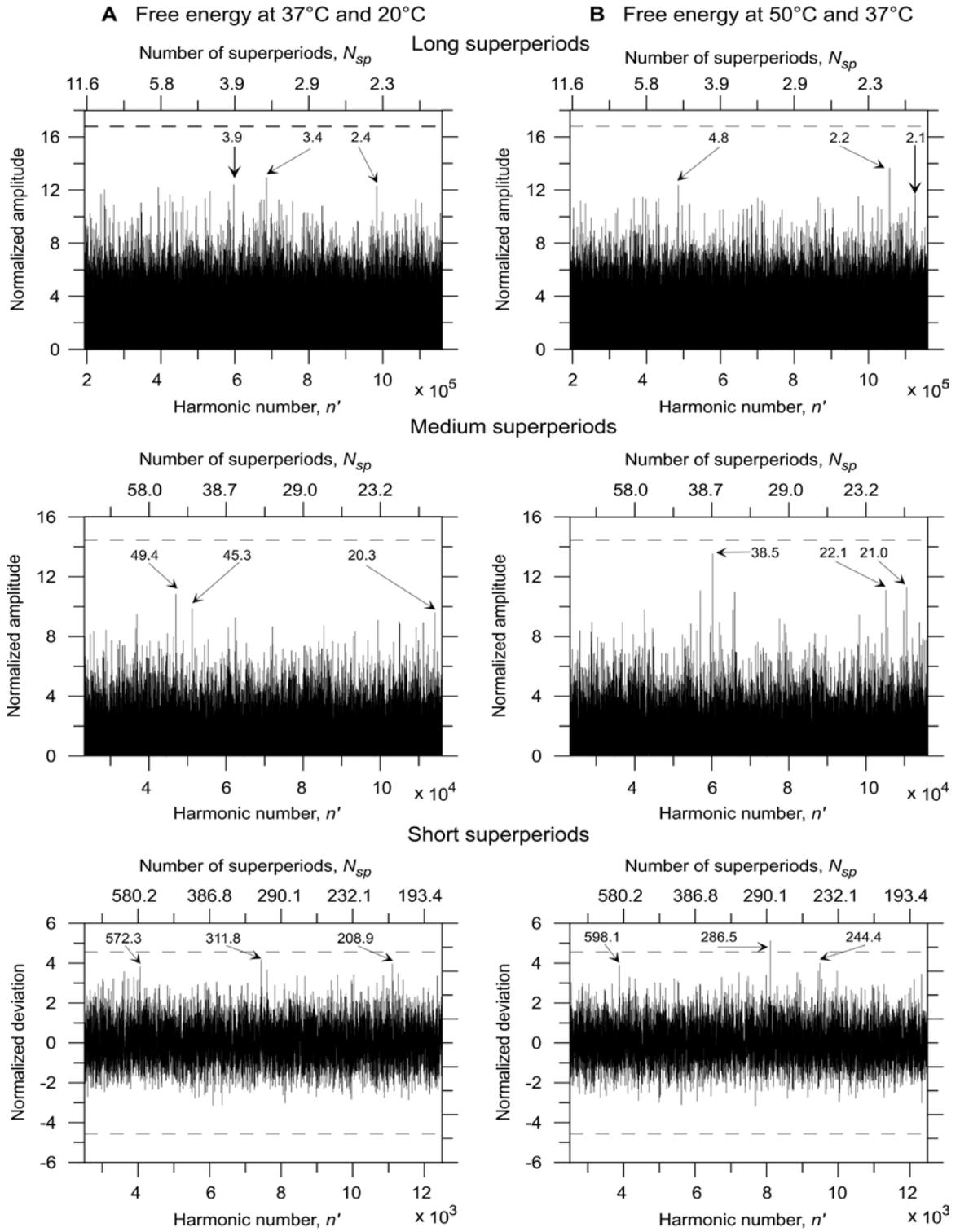

**Fig. 6.** DDFT spectral ranges for the modulational superperiods corresponding to the difference $\Delta f_{t_1|t_2}(q_n) = f_{t_1}(q_n) - f_{t_2}(q_n)$ (Eq. (47)); (**A**) $t_1 = 37°C$, $t_2 = 20°C$; and (**B**) $t_1 = 50°C$, $t_2 = 37°C$. The three highest harmonics in DDFT spectral ranges are indicated by arrows and the numbers of corresponding modulational superperiods are shown explicitly. The horizontal broken lines refer to 5% significance thresholds for extreme value statistics for the chosen spectral ranges.

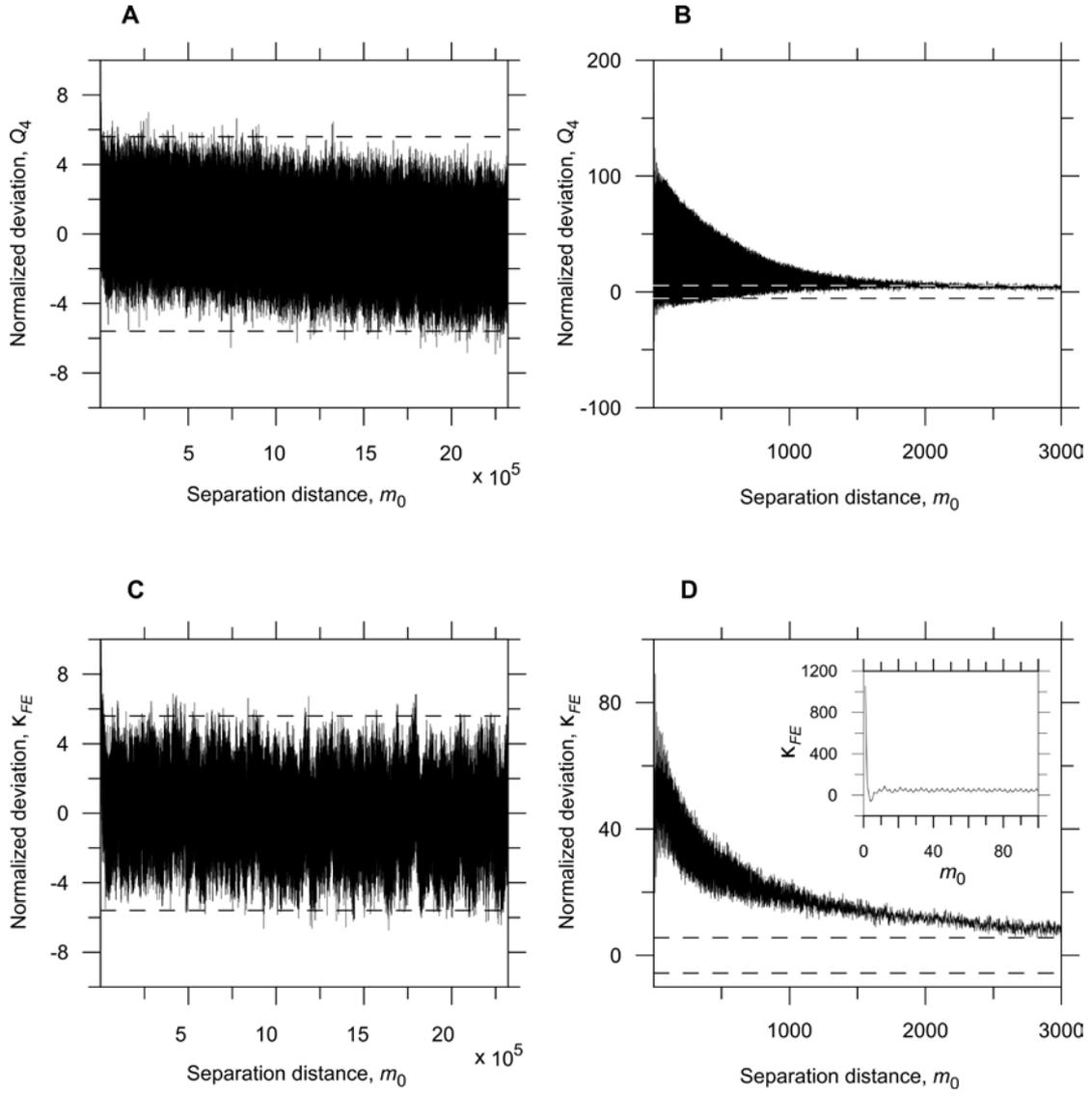

**Fig. 7.** The dependences on the separation distance $m_0$ for the sum of normalized deviations for nucleotide correlations, $Q_4(m_0)$ (Eq. (18)) (**A**), and for the base-pairing free energy profile at 37°C, $\kappa_{FE}(m_0) = (K_{FE}(m_0) - \overline{K}_{FE}) / <\Delta K_{FE}^2 >_{\text{random}}^{1/2}$ (Eqs. (29)–(32)) (**C**). The deviations corresponding to the shorter separation distances $m_0 \leq 3$ kb are shown in details apart (**B** and **D**, respectively). Additionally, the normalized deviations $\kappa_{FE}(m_0)$ at the shortest separation distances $m_0 \leq 100$ bp are shown in the insert to the panel **D**. The horizontal broken lines refer to 5% significance threshold for the Gaussian extreme value statistics in the range $m_0 \leq M/2$, where $M$ is the genome length (Section 2.5).

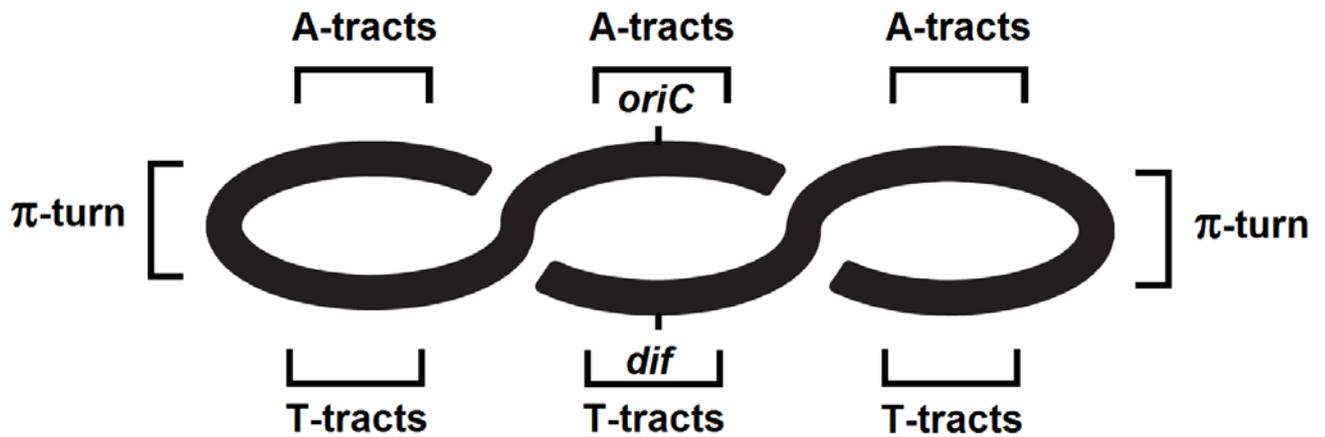

**Fig. 8.** The schematic representation of the *E. coli* chromosome in rapidly growing bacteria. The regions of chromosome enriched by stretches with A-tracts (short sequences $A_n$, $n = 2–8$, phased with the helix pitch) can induce the intrinsic curvature of dsDNA and facilitate the chromosome folding. The T-tracts (short sequences $T_n$, $n = 2–8$, phased with the helix pitch) can induce the reverse intrinsic curvature of dsDNA. The locus *oriC* refers to the origin of chromosome replication, while *dif* refers to the site, where replicated chromosome dimers are resolved.

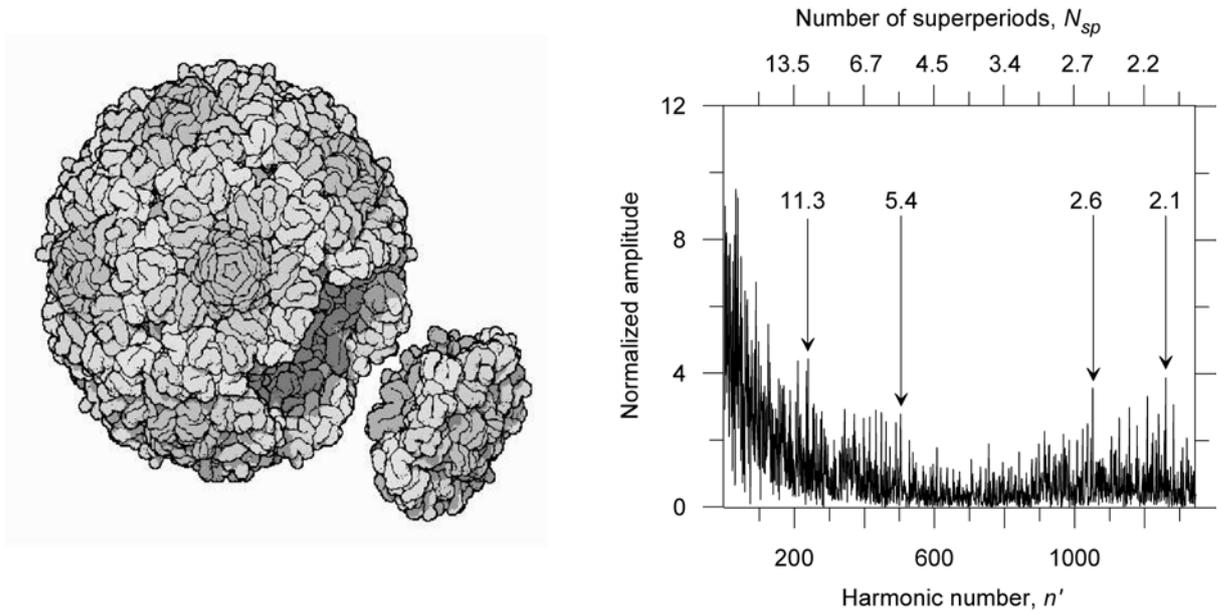

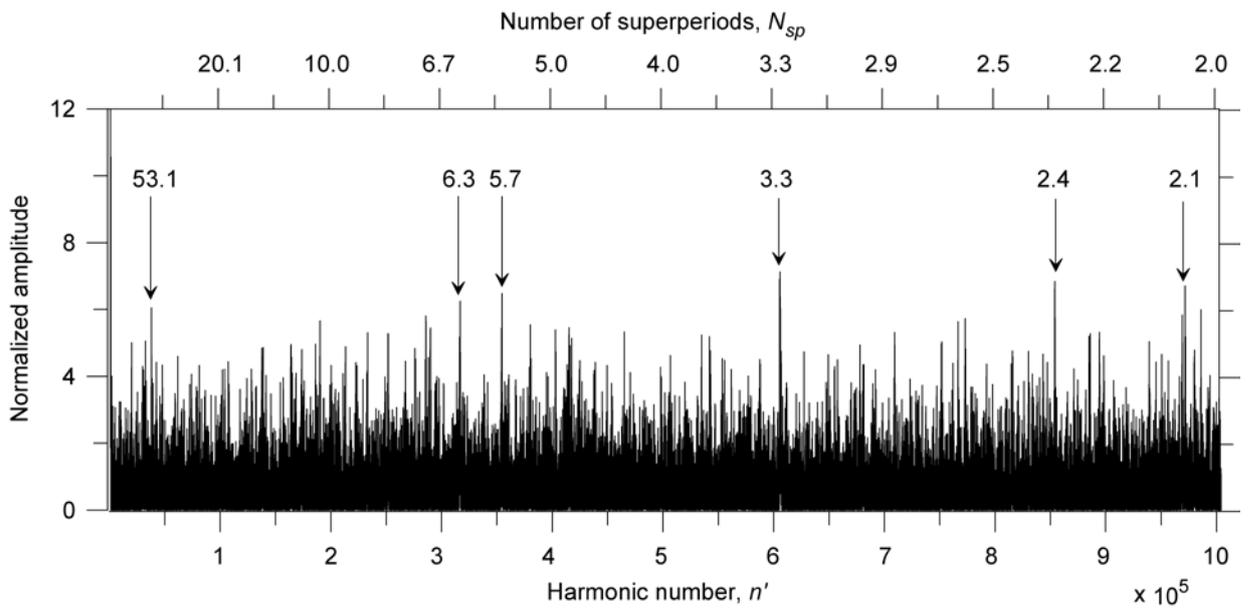

**Fig. 9.** DDFT spectra for the genomes of bacteriophage PHIX174 (**A**, on the right) and bacterium *C. crescentus* CB15 (wild-type) (**B**). The spectra correspond to DDFT of the sum $S_4(q_n)$ (Eq. (8)). The de-trending to both spectra was not applied. The dodecahedron capsid with scaffolding proteins for bacteriophage PHIX174 is shown on the left (**A**). The superperiods corresponding to the highest harmonics in DDFT spectra are marked by arrows.